\documentstyle[preprint,epsf]{jpsj}

\input defphys.inc

\def\runtitle{
Numerical Study of the Ground State
of Multi-Orbital Hubbard Models
}
\def\runauthor{Yukitoshi {\sc Motome} and Masatoshi {\sc Imada}}

\def\sf{\rm}

\title{
Numerical Study for the Ground State \\
of Multi-Orbital Hubbard Models
}

\author{Yukitoshi {\sc Motome}$^{1}$ and Masatoshi {\sc Imada}$^{2}$
}
\inst{
$^{1}$Department of Physics, Tokyo Institute of Technology, \\
Oh-okayama 2-12-1, Meguro-ku, Tokyo 152-8551 \\
$^{2}$Institute for Solid State Physics, University of Tokyo, \\
Roppongi 7-22-1, Minato-ku, Tokyo 106-8666
}
\recdate{
May 11, 1998
}

\abst{
Ground state properties of multi-orbital Hubbard models are
investigated by the auxiliary field quantum Monte Carlo method.
A Monte Carlo technique generalized
to the multi-orbital systems is introduced and examined in detail.
The algorithm contains non-trivial cases where
the negative sign problem does not exist.
We investigate one-dimensional systems
with doubly degenerate orbitals by this new technique.
Properties of the Mott insulating state are quantitatively
clarified as the strongly correlated insulator,
where the charge gap amplitude is much larger than the spin gap.
The insulator-metal transitions driven
by the chemical potential shows a universality class
with the correlation length exponent $\nu=1/2$,
which is consistent with the scaling arguments.
Increasing level split between two orbitals drives crossover
from the Mott insulator with high spin state to
the band insulator with low spin state,
where the spin gap amplitude increases and becomes closer
to the charge gap.
Experimental relevance of our results
especially to Haldane materials is discussed.
}

\kword{
multi-orbital Hubbard model,
orbital degrees of freedom,
auxiliary field quantum Monte Carlo method,
metal-insulator transition,
filling control,
scaling theory,
level split,
Mott insulator, band insulator,
high spin, low spin,
spin gap, Haldane systems
}

\begin{document}
\sloppy
\maketitle

\section{Introduction}
\label{Sec:Introduction}

In the last dozen years since the discovery of high $T_{\rm c}$
materials \cite{Bednorz1986},
$d$ and $f$ electron compounds
have been studied extensively with revived interest
for the purpose of clarifying effects of strong correlations
\cite{Imada1998}.
In these systems,
competition between spin and orbital components
plays an important role leading to complex phenomena
such as metal-insulator transitions
with spin and/or orbital orderings and fluctuations.
Especially, the relevance of orbital degrees of freedom
is a subject of recent intensive studies.
%Progress on sample preparation and experimental technique
%have revealed various critical properties in detail.

The multi-orbital Hubbard model is one of 
the realistic and simplified models
to investigate the interplay between spin and orbital components
in the Mott transition.
This model is straightforwardly derived
from the tight-binding scheme.
The explicit form of the Hamiltonian is given by
\begin{equation}
\label{generalH}
{\cal H} = 
{\cal H}_{t} + {\cal H}_{\varepsilon} +
{\cal H}_{\mu} + {\cal H}_{U} + {\cal H}_{J},
\end{equation}
where
\begin{eqnarray}
\label{Ht}
{\cal H}_{t} &=& 
\sum_{ij} \sum_{\nu\nu'} t_{ij}^{\nu\nu'}
\left( c_{i\nu}^{\dagger} c_{j\nu'} + {\rm h.c.} \right)
\\
{\cal H}_{\varepsilon} &=& 
\sum_{i} \sum_{\nu} \varepsilon_{\nu} n_{i\nu}
\\
{\cal H}_{\mu} &=& - \mu \sum_{i} \sum_{\nu} n_{i\nu}
\\
\label{HU}
{\cal H}_{U} &=& \sum_{i} \sum_{\nu\leq\nu'} U_{\nu\nu'}
                 \sum_{\sigma\leq\sigma'}
                 \left(1-\delta_{\nu\nu'}\delta_{\sigma\sigma'}\right)
                 n_{i\nu\sigma} n_{i\nu'\sigma'}
\\
\label{HJ}
{\cal H}_{J} &=& \sum_{i} \sum_{\nu\neq\nu'}
%                  \frac{J_{\nu\nu'}}{2} \sum_{\sigma\sigma'}
                  J_{\nu\nu'} \sum_{\sigma\leq\sigma'}
                 \bigl( c_{i\nu\sigma}^{\dagger} c_{i\nu'\sigma'}^{\dagger}
                        c_{i\nu\sigma'} c_{i\nu'\sigma}
%\\ \nonumber & & \qquad \qquad \qquad \qquad
                      + c_{i\nu\sigma}^{\dagger} c_{i\nu\sigma'}^{\dagger}
                        c_{i\nu'\sigma'} c_{i\nu'\sigma} \bigr).
\end{eqnarray}
Here the operator
$c_{i\nu}^{\dagger}$($c_{i\nu}$) creates(annihilates) an electron
in the orbital $\nu = 1,\cdot\cdot\cdot,N_{\rm C}$
on the site $i = 1,\cdot\cdot\cdot,N_{\rm S}$, and
$n_{i\nu} \equiv c_{i\nu}^{\dagger} c_{i\nu}$
defines the number operator.
The first three terms are one-body terms:
${\cal H}_{t}$ is the hopping term,
${\cal H}_{\varepsilon}$ gives the relative energy level
for each orbital and
${\cal H}_{\mu}$ sets the averaged chemical potential $\mu$
which controls the electron filling;
We take $\sum_{\nu}\varepsilon_{\nu}=0$.
The last two,
${\cal H}_{U}$ and ${\cal H}_{J}$, are two-body interaction terms
which are classified as in the conventional way:
The former term ${\cal H}_{U}$ is often called the on-site Coulomb term
which consists of only diagonal interactions.
The latter ${\cal H}_{J}$ is often called the on-site exchange term
which contains
the Hund coupling and the pair-hopping between orbitals.
Note that,
in order to keep the rotational symmetry of the two-body interaction,
the parameters in Eqs. (\ref{HU}) and (\ref{HJ}) are
not independent of each other originally
\cite{Brandow1977,Kanamori1963}.
For example, for the twofold $e_{g}$ orbitals in $d$ electron systems
whose wave functions are given by
$\psi_{i1\sigma} = \left(x^{2}-y^{2}\right) f\left(r\right)$ and
$\psi_{i2\sigma} = \frac{1}{\sqrt{3}} \left(3z^{2}-r^{2}\right)
f\left(r\right)$,
the relation $U_{11} = U_{22} = U_{12} + 2J_{12}$
should be satisfied.
%In the following parts,
%we assume that these parameters are independent of each other
%for theoretical interest.

As mentioned above, in these multi-orbital systems,
fluctuations of both spin and orbital degrees of freedom
may compete each other.
Various approximations have been
examined on this model Eq. (\ref{generalH});
mean-field approximations
\cite{Roth1966a,Roth1966b,Kugel1982,Kugel1973,Inagaki1973,Cyrot1975},
Gutzwiller approximations
\cite{Bunemann1997,Okabe1997,Okabe1996,Chao1971a,Chao1971b,Lu1994},
infinite-dimensional approaches
\cite{Kotliar1996,Rozenberg1997}
and slave boson approaches
\cite{Hasegawa1997a,Hasegawa1997b,Hasegawa1997c}.
However, these approximations in general neglect fluctuations
under the competition of various orderings.
Although details of the mean-field approximations
in these approaches differ each other,
they all have a difficulty
in describing critical phenomena by nature.
A more appropriate treatment
which allows studies on fluctuation effects has to be employed
in order to discuss critical nature of phase transitions
in these complex systems.
In fact, the criticality may have relevance
to many aspects of physical properties in $d$ or $f$ electron compounds.

In this work,
we investigate ground state properties of
the multi-orbital Hubbard model
by an unbiased method, that is,
by the auxiliary field quantum Monte Carlo (QMC) method,
recently proposed by the authors \cite{Motome1997}.
Several remarkable aspects which cannot be obtained
within one-body approximations are elucidated
for the one-dimensional systems with doubly degenerate orbitals:
First, the characteristic properties as the strongly correlated insulator
are elucidated quantitatively for the Mott insulator at half filling;
though the charge and spin excitations both have gaps,
the charge gap amplitude is much larger than the spin gap.
The orbital excitation is gapless, that is,
the orbital polarization grows continuously
with increasing the level split.
Critical properties of the insulator-metal transition
are studied by controlling the chemical potential.
Analysis on the numerical results based on the hyperscaling hypothesis
gives the correlation length exponent $\nu=1/2$,
which supports the scaling arguments
on the Mott transition in one dimension.
The properties under the self doping
between two orbitals are also investigated
by controlling the level split;
the continuous change from the high-spin Mott insulator
to the low-spin band insulator is elucidated,
where the spin gap amplitude increases
sensitively depending on the magnitude of the self doping
and become closer to the charge gap.

This paper is organized as follows:
In the next section,
the auxiliary field QMC technique proposed by the authors
is reviewed in detail in a general formalism
for arbitrary number of orbital in arbitrary dimension.
The matrix representation for this formalism and
details of QMC updating are also described explicitly
for the readers who are interested in the actual coding procedure.
We show that the algorithm does not suffer from the negative sign problem
in a non-trivial and physically interesting parameter region
of the model defined in Eq. (\ref{generalH}).
In Sec. \ref{Sec:DDM}, we introduce
doubly degenerate models to investigate them in the following sections.
We also discuss experimental relevance
of this doubly degenerate model.
In Sec. \ref{Sec:Results}, the ground state properties of this model
are investigated by the QMC method in one dimension.
Properties of the Mott insulating state at half filling
is investigated in terms of correlations
of charge, spin and orbital degrees of freedom.
The critical properties of insulator-metal transitions
controlled by the chemical potential are discussed.
For finite level split between two orbitals,
effects of self doping are investigated.
All these QMC results are discussed in Sec. \ref{Sec:Discussion}
in detail.
Based on the scaling analysis,
the critical exponents of the insulator-metal transition
by filling control are discussed.
The crossover between the Mott insulator in high spin state
and the band insulator in low spin state is examined
for the level-split control.
Finally, Sec. \ref{Sec:Summary} is devoted to summary.

\section{Method}
\label{Sec:Method}

Auxiliary field quantum Monte Carlo method is a powerful tool
to investigate interacting fermions
\cite{Hirsch1983,Hirsch1985,Sugiyama1986,White1989,Imada1989}.
In this section,
we describe a general algorithm
for multi-orbital Hubbard models defined in Eq. (\ref{generalH})
for arbitrary number of orbital in arbitrary dimension.
This algorithm was developed in the previous letter
by the authors \cite{Motome1997}.
We describe it in detail
to make this paper be self-contained,
especially supplementing the matrix representation
and the details of updating procedure.

%%%%%%%%%%
\subsection{Assumption}
\label{Sec:Assumption}

For the efficiency of the algorithm described below,
we assume $U_{\nu\nu'} = U$ in Eq. (\ref{HU}).
This allows us to factorize the two-body interaction terms
Eqs. (\ref{HU}) and (\ref{HJ})
into the quadratic forms, as
\begin{eqnarray}
\label{HUquad} 
{\cal H}_{U} &=& \frac{U}{2} \sum_{i}
\left(n_{i}-N_{\rm C}\right)^{2}
%\\ \nonumber
%&+&
+
\frac{U}{2} \left(2N_{\rm C}-1\right) \sum_{i} n_{i}
- \frac{U}{2} N_{\rm C}^{2}
\\
\label{HJquad}
{\cal H}_{J} &=& \sum_{i} \sum_{\nu<\nu'} 
                 \frac{J_{\nu\nu'}}{2} A_{i\nu\nu'}^{2}
%\\ \nonumber
%&-&
-
%\Bigl(N_{\rm C}-1\Bigr)
                \sum_{i} \sum_{\nu<\nu'} 
                 \frac{J_{\nu\nu'}}{2} n_{i}.
\end{eqnarray}
Here
$A_{i\nu\nu'} \equiv \sum_{\sigma} A_{i\nu\nu'\sigma}
\equiv \sum_{\sigma} 
( c_{i\nu\sigma}^{\dagger} c_{i\nu'\sigma}
+ c_{i\nu'\sigma}^{\dagger} c_{i\nu\sigma} )$.
These factorizations simplify
the Hubbard-Stratonovich (HS) transformation
as is described below
in Eqs. (\ref{expU}) and (\ref{expJ}).

As mentioned in the introduction,
the interaction parameters $U_{\nu\nu'}$ and $J_{\nu\nu'}$
depend on each other to keep the rotational symmetry
of the total interaction
\cite{Brandow1977,Kanamori1963}.
Our assumption $U_{\nu\nu'} = U$ breaks this symmetry.
%%%(See Eq. (\ref{spisprep}) for the doubly degenerate case.)
%We also note that there exist particular cases where
%this assumption may introduce
%an additional degeneracy in the atomic limit $U \gg t$
%depending on the electron filling and the numbers of orbital degeneracy.
%For example, in the case with one electron per site
%in doubly degenerate orbitals,
%the spin-ferro and orbital-antiferro state degenerates
%the spin-antiferro and orbital-ferro state
%in the second perturbation in $t/U$ from the atomic limit.
In the following, we further treat 
the values of $U$ and $J_{\nu\nu'}$ as independent parameters
as in many previous theoretical studies.
We believe that 
these may not affect the qualitative features of the problem.
Note that although our assumptions break the on-site rotational symmetry,
%%%especially of the orbital part
the symmetry or dimension of elementary excitations might be
closely related to the effective exchange coupling
between different sites
which strongly depends on the specific form of
the hopping matrix $t_{ij}^{\nu\nu'}$
as well as the electron filling or the orbital degeneracy;
%%%\cite{comment} ;
the following algorithm is applicable to any type of $t_{ij}^{\nu\nu'}$.
%We at least expect that
%even under these restrictions the model Eq. (\ref{generalH})
%allows us to clarify basic properties
%as far as the critical properties and the universality are concerned.

%%%%%%%%%%
\subsection{Basic formalism}
\label{BF}

There exist two schemes in the auxiliary field QMC technique;
one is for finite temperatures and
the other is for the ground state.
In the finite temperature calculation,
we need to calculate the partition function,
$Z={\rm Tr} \exp \left(-\beta {\cal H} \right)$
\cite{Hirsch1983,Hirsch1985,Sugiyama1986}.
Compared with this, for the ground state calculation
which is often called the projection quantum Monte Carlo method
\cite{White1989,Imada1989},
it is necessary to calculate the density matrix,
$\rho \left( \beta ; \phi \right) \equiv
\langle \phi | \exp \left( -\beta {\cal H} \right) | \phi \rangle$,
where $| \phi \rangle$ is a trial wave function
non-orthogonal to the ground state.
Below,
we describe the basic formalism to calculate the common quantity
$\exp \left( -\beta {\cal H} \right)$
in both schemes.

First step is the Suzuki-Trotter decomposition into
the $M = \beta / \Delta\tau$ slices
in the imaginary time direction 
\cite{Trotter1959,Suzuki1976a,Suzuki1976b}.
Then we obtain
%\begin{eqnarray}
\begin{equation}
\label{STdecomp}
e^{-\beta{\cal H}}
%&=&
=
\left(
e^{-\Delta\tau {\cal H}_{t}/2}
e^{-\Delta\tau {\cal H}_{\varepsilon}/2}
e^{-\Delta\tau {\cal H}_{\mu}}
%\right. \\ \nonumber
%& & \left.
e^{-\Delta\tau {\cal H}_{U}}
e^{-\Delta\tau {\cal H}_{J}}
e^{-\Delta\tau {\cal H}_{\varepsilon}/2}
e^{-\Delta\tau {\cal H}_{t}/2}
\right)^{M}
%\\ \nonumber
%& &
+ O\left( \Delta\tau^{2} \right).
%\end{eqnarray}
\end{equation}
Next, in order to derive a one-body expression,
we replace the two-body interaction terms
${\cal H}_{U}$ and ${\cal H}_{J}$
with non-interacting ones
by introducing the HS transformation
\cite{Hubbard1959,Stratonovich1957}.
The general formula of the discrete HS transformation
we use here is given by \cite{Motome1997}
\begin{equation}
\label{HSgeneral}
\exp\left(-\Delta\tau \theta f^{2}\right)
= \sum_{l,s=\pm1} \frac{\gamma_{l}}{4}
\exp\left( {\rm i} s \eta_{l} \sqrt{\alpha} f \right) 
+ O(\Delta\tau^{4}),
\end{equation}
where
$\gamma_{l} \equiv 1+\frac{\sqrt{6}}{3}l,
\eta_{l} \equiv \sqrt{2(3-\sqrt{6}l)}$
and $\alpha \equiv \Delta\tau \theta \geq 0$.
Applying this formula to Eqs. (\ref{HUquad}) and (\ref{HJquad}),
we obtain
\begin{eqnarray}
\label{expU} 
& &
\exp \Bigl( -\Delta\tau 
\frac{U}{2} \sum_{i} \left( n_{i}-N_{\rm C}\right)^{2} \Bigr)
\\ \nonumber
&=& \prod_{i} \biggl[ \sum_{l_{1i},s_{1i}=\pm1}
\frac{\gamma_{l_{1i}}}{4} \exp \left\{ {\rm i} s_{1i} \eta_{l_{1i}}
\sqrt{\alpha_{1}} \left( n_{i}-N_{\rm C} \right) \right\}\biggr]
%\\
%& &
 + O(\Delta\tau^{4})
\nonumber
 \\
\label{expJ}
& &
 \exp \Bigl( -\Delta\tau 
\sum_{i}\sum_{\nu<\nu'} \frac{J_{\nu\nu'}}{2} 
A_{i\nu\nu'}^{2} \Bigr)
\\ \nonumber
&=& \prod_{i} \prod_{\nu<\nu'} \biggl[ 
\sum_{l_{2i}^{\nu\nu'},s_{2i}^{\nu\nu'}=\pm1}
\frac{\gamma_{l_{2i}^{\nu\nu'}}}{4} 
%\\ \nonumber
%& & \qquad \qquad
\exp \left\{ {\rm i} s_{2i}^{\nu\nu'} \eta_{l_{2i}^{\nu\nu'}}
\sqrt{\alpha_{2}^{\nu\nu'}} A_{i\nu\nu'} \right\} \biggr]
+ O(\Delta\tau^{4}),
\nonumber
\end{eqnarray}
where $\alpha_{1}=\Delta\tau U/2 \ge 0$ and 
$\alpha_{2}^{\nu\nu'}=\Delta\tau J_{\nu\nu'}/2 \ge 0$.

Of course, the way to decompose the two-body interaction terms,
that is, the type of the HS transformation is not unique.
The advantage of our scheme with Eq. (\ref{HSgeneral})
lies in the fact that it brings the non-trivial parameter regions
where the negative sign difficulty does not exist.
We discuss this point in Sec. \ref{Sec:Conditions} in detail.

We apply these decompositions to each Suzuki-Trotter slice
in Eq. (\ref{STdecomp}).
Then, the one-body expression is given as
%\begin{eqnarray}
\begin{equation}
%& &
 \exp \left(-\beta {\cal H} \right)
%\nonumber \\
%&=&
=
\sum_{\{l_{1}s_{1}l_{2}^{\nu\nu'}s_{2}^{\nu\nu'}\}}
\prod_{\sigma} \prod_{m=1}^{M} \left[
w_{t\sigma} w_{\varepsilon\sigma} w_{\mu\sigma}
%\right. \\ \nonumber
%& & \left. 
w_{U\sigma}(l_{1}(m),s_{1}(m))
w_{J\sigma}(l_{2}(m),s_{2}(m)) 
w_{\varepsilon\sigma} w_{t\sigma}
\right]
%\end{eqnarray}
\end{equation}
with
\begin{eqnarray}
w_{t\sigma} &=& \prod_{i=1}^{N_{\rm S}} 
\exp\left(-\Delta\tau{\cal H}_{t\sigma}/2\right) \\
w_{\varepsilon\sigma} &=& \prod_{i=1}^{N_{\rm S}} \prod_{\nu}
\exp\left( -\Delta\tau \varepsilon_{\nu} n_{i\nu\sigma}/2 \right)\\
w_{\mu\sigma} &=& \left[ -\mu 
                         + \frac{U}{2}\left(2N_{\rm C}-1\right)
                         + \sum_{\nu<\nu'}\frac{J_{\nu\nu'}}{2}
                       \right]
                  \sum_{i} n_{i\sigma} \\
w_{U\sigma} &=& \prod_{i=1}^{N_{\rm S}}
%\left[
\frac{\sqrt{\gamma_{l_{1i}}}}{2}
%\right. \\ \nonumber
%& & \qquad \qquad \left.
\exp \Bigl\{ {\rm i} s_{1i}\eta_{l_{1i}}
\sqrt{\alpha_{1}} \Bigl(n_{i\sigma}-\frac{N_{\rm C}}{2}
\Bigr) \Bigr\}
%\right]
\\
w_{J\sigma} &=& \prod_{{i=1}\atop{\nu<\nu'}}^{N_{\rm S}}
%\prod_{\nu<\nu'}
%\left[
\frac{\sqrt{\gamma_{l_{2i}^{\nu\nu'}}}}{2}
%\right. \\ \nonumber
%& & \qquad \qquad \left.
\exp \left( {\rm i} s_{2i}^{\nu\nu'}\eta_{l_{2i}^{\nu\nu'}}
\sqrt{\alpha_{2}^{\nu\nu'}} A_{i\nu\nu'\sigma} \right) .
%\right].
\end{eqnarray}
The product of Eqs. (\ref{expU}) and (\ref{expJ})
over all the Suzuki-Trotter slices amounts to 
the systematic error of $O(\Delta\tau^{3})$
which is negligible because it is higher order than
the other systematic errors 
from the Suzuki-Trotter decomposition in Eq. (\ref{STdecomp}).
In Appendix A, we discuss the behavior of these errors in detail.

%%%%%%%%%%
\subsection{Matrix representation}
\label{Sec:MR}

A matrix representation for the above expressions
is helpful in an actual coding.
We derive it explicitly here in the case of the ground state algorithm
for discussions in the following part.
The following representations can easily be
applied to the finite temperature calculation
only with slight modifications.

We use a trial wave function
$| \phi \rangle = | \phi^{\uparrow} \rangle 
\otimes | \phi^{\downarrow} \rangle$
with
\begin{equation}
| \phi^{\sigma} \rangle =
\prod_{j=1}^{N_{\rm e}^{\sigma}} \left(
\sum_{i=1}^{N} \left( \Phi_{\sigma} \right)_{ij} c_{i\sigma}^{\dagger}
\right) | 0 \rangle,
\end{equation}
where
$N_{\rm e}^{\sigma}$ is the electron number with $\sigma$ spin;
$N = N_{\rm S} \times N_{\rm C}$; and
$| 0 \rangle$ is the vacuum.
Then, the matrix representation for the density matrix 
in the canonical calculation is given as
%\begin{eqnarray}
\begin{equation}
\label{dmW}
\rho\left(\beta;\phi\right)
%&=&
=
\sum_{\{l_{1}s_{1}l_{2}^{\nu\nu'}s_{2}^{\nu\nu'}\}}
W_{\uparrow}\left(\{l_{1}s_{1}l_{2}^{\nu\nu'}s_{2}^{\nu\nu'}\};
\beta;\phi\right)
%\\ \nonumber
%& & \qquad \qquad
W_{\downarrow}\left(\{l_{1}s_{1}l_{2}^{\nu\nu'}s_{2}^{\nu\nu'}\};
\beta;\phi\right),
%\end{eqnarray}
\end{equation}
where
\begin{equation}
W_{\sigma} =
C_{\{l_{1}s_{1}l_{2}^{\nu\nu'}s_{2}^{\nu\nu'}\}}
\det \left[ ^{t}\Phi_{\sigma}
B_{\{l_{1}s_{1}l_{2}^{\nu\nu'}s_{2}^{\nu\nu'}\}}
\left( \beta,0 \right)
\Phi_{\sigma} \right].
\end{equation}
Here,
%\begin{eqnarray}
\begin{equation}
C_{\{l_{1}s_{1}l_{2}^{\nu\nu'}s_{2}^{\nu\nu'}\}} 
%&=&
=
\prod_{m=1}^{M} \prod_{i=1}^{N}
\frac{\sqrt{\gamma_{l_{1i}\left(m\right)}
\gamma_{l_{2i}^{\nu_{i}{\nu_{i}}'}\left(m\right)}}}{4} 
%\\ \nonumber
%& &
\exp \left( -{\rm i} s_{1i}\left(m\right) \eta_{l_{1i}\left(m\right)}
\sqrt{\alpha_{1}} \frac{N_{C}}{2} \right)
%\end{eqnarray}
\end{equation}
\begin{equation}
B_{\{l_{1}s_{1}l_{2}^{\nu\nu'}s_{2}^{\nu\nu'}\}}
\left( \tau_{2},\tau_{1} \right) =
\prod_{m=m_{1}+1}^{m_{2}} b \left(m\right),
\end{equation}
where $\tau_{i} = m_{i} \Delta\tau$ $(i=1,2)$ and
\begin{equation}
b \left(m\right) =
e^{-\frac{\Delta\tau}{2} T} e^{L}
e^{D(m)} e^{E(m)} e^{L} e^{-\frac{\Delta\tau}{2} T}.
\end{equation}
Here
$T,L,D$ and $E$ are $N \times N$ matrices given by
\begin{eqnarray}
\label{MRHt}
\sum_{\tilde{i},\tilde{j}}^{N} c_{\tilde{i}\sigma}^{\dagger}
\left(T\right)_{\tilde{i}\tilde{j}} c_{\tilde{j}\sigma}
&=& {\cal H}_{t\sigma}
\\
\left(L\right)_{\tilde{i}\tilde{j}} &=& -\frac{\Delta\tau}{2}
\varepsilon_{\nu_{i}} \delta_{\tilde{i}\tilde{j}}
\\
\left(D\right)_{\tilde{i}\tilde{j}} &=& {\rm i} 
s_{1{i}}\left(m\right)
\eta_{l_{1{i}}\left(m\right)}
\sqrt{\alpha_{1}} \delta_{ij}
\\
\label{MRE}
\left(E\right)_{\tilde{i}\tilde{j}} &=& \sum_{\nu<\nu'}
{\rm i} s_{2{i}}^{\nu\nu'}\left(m\right)
\eta_{l_{2{i}}^{\nu\nu'}\left(m\right)}
%\\ \nonumber & &
\sqrt{\alpha_{2}^{\nu\nu'}} \delta_{ij}
\left( \delta_{\nu_{i}\nu} \delta_{\nu_{j}\nu'}
+ \delta_{\nu_{j}\nu} \delta_{\nu_{i}\nu'} \right).
\end{eqnarray}
In Eqs. (\ref{MRHt}) $\sim$ (\ref{MRE}),
the index $\tilde{i}$ contains the site and orbital indices,
$\tilde{i} \equiv \left(i,\nu_{i}\right)$.
$\delta_{ij}$ is the Cronecker's delta function.

%%%%%%%%%%
\subsection{Updating}
\label{Sec:Updating}

The summation over $\{l_{1}s_{1}l_{2}^{\nu\nu'}s_{2}^{\nu\nu'}\}$
in Eq. (\ref{dmW}) is statistically sampled by the Monte Carlo technique.
In the ground state algorithm, a change
$\{l,s\} = \{l_{1}s_{1}l_{2}^{\nu\nu'}s_{2}^{\nu\nu'}\} \rightarrow
\{l',s'\} = \{l'_{1}s'_{1}{l'}_{2}^{\nu\nu'}{s'}_{2}^{\nu\nu'}\}$
is accepted by the probability
\begin{equation}
\label{prob}
P = \left| \frac
{\prod_{\sigma} W_{\sigma} \left( \{l',s'\} \right)}
{\prod_{\sigma} W_{\sigma} \left( \{l,s\} \right)} \right|.
\end{equation}

The probability Eq. (\ref{prob}),
which is called the acceptance ratio, may be rewritten
into more simple form by using the single-particle Green function
\cite{Imada1989}.
For a change from
$\{l_{\lambda i}\left(m\right), s_{\lambda i}\left(m\right)\}$
to
$\{\bar{l}_{\lambda i}\left(m\right), \bar{s}_{\lambda i}\left(m\right)\}$
$(\lambda = 1,2)$,
the acceptance ratio $P$ is given as
\begin{equation}
\label{explicitP}
P = \left| \tilde{C}_{\lambda} \prod_{\sigma} \det \left[
I + \Delta_{\lambda} G_{\sigma} \left(m\right) \right] \right|,
\end{equation}
where $I$ is the $N \times N$ unit matrix,
\begin{eqnarray}
\tilde{C}_{1} &=& 
\sqrt{
\frac{\gamma_{\bar{l}_{1i}\left(m\right)}}{\gamma_{l_{1i}\left(m\right)}}
}
\exp \left[
-{\rm i} \sqrt{\alpha_{1}}\frac{N_{\rm C}}{2}
%\right. \\ \nonumber
%& & \left. \quad
\left( \bar{s}_{1i}\left(m\right) \eta_{\bar{l}_{1i}\left(m\right)}
     - s_{1i}\left(m\right) \eta_{l_{1i}\left(m\right)} \right)
\right]
\\
\tilde{C}_{2} &=&
\sqrt{
\gamma_{\bar{l}_{2i}^{\nu_{i}{\nu_{i}}'}\left(m\right)} /
\gamma_{l_{2i}^{\nu_{i}{\nu_{i}}'}\left(m\right)}
}.
\end{eqnarray}
$\Delta_{\lambda}$ are sparse $N \times N$ matrices with
only non-zero elements,
\begin{equation}
 \left( \Delta_{\lambda} \right)_{\left(i,\nu_{i}\right)
                           \left(i,{\nu_{i}}'\right)}
=
\left( \Gamma_{\lambda} \right)_{\left(i,\nu_{i}\right)
                           \left(i,{\nu_{i}}'\right)}
- \delta_{\nu_{i}{\nu_{i}}'}
\end{equation}
with
\begin{eqnarray}
\left( \Gamma_{1} \right)_{\tilde{i} \tilde{i}}
&=&
\exp \left[
{\rm i} \sqrt{\alpha_{1}}
\left( \bar{s}_{1i}\left(m\right) \eta_{\bar{l}_{1i}\left(m\right)}
     - s_{1i}\left(m\right) \eta_{l_{1i}\left(m\right)} \right)
\right]
\\
\left( \Gamma_{2} \right)_{\left(i,\nu_{i}\right)
                           \left(i,{\nu_{i}}'\right)}
&=&
\frac{
\exp \left[ {\rm i} \sqrt{\alpha_{2}^{\nu_{i}{\nu_{i}}'}}
\bar{s}_{2i}^{\nu_{i}{\nu_{i}}'}\left(m\right)
\eta_{\bar{l}_{2i}^{\nu_{i}{\nu_{i}}'}\left(m\right)} \right]
}
{
\exp \left[ {\rm i}\sqrt{\alpha_{2}^{\nu_{i}{\nu_{i}}'}}
s_{2i}^{\nu_{i}{\nu_{i}}'}\left(m\right)
\eta_{l_{2i}^{\nu_{i}{\nu_{i}}'}\left(m\right)} \right]
}.
\end{eqnarray}
Here, the single-particle Green function
at the $m$-th imaginary time slice is calculated by
\begin{equation}
G_{\sigma}\left(m\right) = \Theta_{R}^{\sigma}\left(m\right)
\left( ^{t}\Phi_{\sigma} B\left(\beta,0\right) \Phi_{\sigma} \right)^{-1}
 { }^{t}\Theta_{L}^{\sigma}\left(m\right),
\end{equation}
where
\begin{eqnarray}
\Theta_{R}^{\sigma}\left(m\right) &=&
e^{D\left(m\right)} e^{E\left(m\right)} e^{L}
e^{-\Delta\tau T/ 2}
%\\ \nonumber
%& & \qquad
B\left( \left(m-1\right)\Delta\tau, 0\right) \Phi_{\sigma}
\\
 ^{t}\Theta_{L}^{\sigma}\left(m\right) &=&
 ^{t}\Phi_{\sigma} 
B \left(\beta, \left(m+1\right)\Delta\tau \right) 
e^{-\Delta\tau T / 2} e^{L}.
\end{eqnarray}
Note that
$W_{\sigma} = C \det \left[
 ^{t}\Theta_{L}^{\sigma} \Theta_{R}^{\sigma} \right]$.

If the change of the HS variables is accepted,
the updated Green function $\bar{G}$ for 
$\{\bar{l}_{\lambda i}\left(m\right), \bar{s}_{\lambda i}\left(m\right)\}$
is calculated from the old one $G$ for
$\{l_{\lambda i}\left(m\right), s_{\lambda i}\left(m\right)\}$
as,
\begin{equation}
\label{updateG}
\bar{G}_{\sigma} = \left(I + \Delta_{\lambda}\right)
\tilde{G}_{\sigma},
\end{equation}
where
\begin{eqnarray}
\left(\tilde{G}_{\sigma}\right)_{\tilde{j}\tilde{k}}
&=& \left(G_{\sigma}\right)_{\tilde{j}\tilde{k}}
%\\ \nonumber
%&-&
-
 \sum_{\nu_{i}{\nu_{i}}'}
\left(G_{\sigma}\right)_{\tilde{j}\left(i,\nu_{i}\right)}
\left( \Gamma_{\lambda} \right)_{\nu_{i} {\nu_{i}}'}
\sum_{\nu}
\left(Q^{-1}\right)_{{\nu_{i}}' \nu}
\left(G_{\sigma}\right)_{\left(i,\nu \right)\tilde{k}} .
\end{eqnarray}
The $N_{\rm C} \times N_{\rm C}$ matrix $Q$ is defined as
\begin{eqnarray}
\left(Q\right)_{\nu_{i}{\nu_{i}}'} &=& \delta_{\nu_{i}{\nu_{i}}'}
%\\ \nonumber
%&+&
+
 \sum_{\nu}
\left(G_{\sigma}\right)_{\left(i,\nu_{i}\right)
                         \left(i,\nu\right)}
\left(\Delta_{\lambda}\right)_{\left(i,\nu\right)
                               \left(i,{\nu_{i}}'\right)} .
\end{eqnarray}

To summarize, the actual MC sampling goes as follows:
Calculate the single-particle Green function for given HS variables.
Generate a change of a HS variable and
calculate the acceptance ratio Eq. (\ref{explicitP})
by using the Green function.
If the change is accepted,
recalculate the Green function by Eq. (\ref{updateG}).
Repeat this procedure for all the sites and
all the imaginary time slices.
This amounts to the one MC sweep.

In general,
the weight for each Monte Carlo sample $W_{\uparrow} W_{\downarrow}$
can have a negative value,
which leads to the negative sign problem
\cite{Loh1990,Furukawa1991b,Batrouni1990b,Nakamura1992,Sorella1988,Sorella1989,Fahy1990,Hamman1990,Fahy1991}.
The difficulty caused by the negative sign
depends on details of the systems;
the long-range hopping, the electron density
and the dimensionality of the system.
In Appendix B, we discuss how the negative-sign samples appear 
in our framework
briefly for a specific model which is introduced in Sec. \ref{DDM}.

%%%%%%%%%%
\subsection{Cases free from the negative sign problem}
\label{Sec:Conditions}

Here we discuss non-trivial conditions
on parameters in multi-orbital models
which are completely free from the negative sign problem
\cite{Motome1997}.

Negative-sign samples do not appear
when a system keeps the particle-hole symmetry.
There, it is easily shown that
the weight of the Monte Carlo sample for the up-spin
is just the complex conjugate of that for the down-spin;
$W_{\uparrow} W_{\downarrow} = |W_{\uparrow}|^{2} \ge 0$.
In multi-orbital cases,
the symmetry about orbital indices provides us with
the following conditions i $\sim$ iv
to hold the particle-hole symmetry:

i) The system is at half filling; the electron density satisfies
\begin{equation}
\label{nosign1}
n \equiv \frac{N_{\rm e}^{\uparrow} 
+ N_{\rm e}^{\downarrow}}{N_{\rm S}} = N_{\rm C}.
\end{equation}

ii) The energy levels split symmetrically around the Fermi energy;
\begin{equation}
\label{nosign2}
\varepsilon_{\nu} = - \varepsilon_{\bar{\nu}},
\end{equation}
where ${\bar \nu} \equiv N_{\rm C}-\nu+1$.

iii) The intra-site exchange couplings satisfy a particular relation;
\begin{equation}
\label{nosign3}
J_{\nu\nu'} = \delta_{{\bar \nu}\nu'} J_{\nu}.
\end{equation}

iv) The hopping integrals satisfy a special condition;
\begin{equation}
\label{nosign4}
t_{ij}^{\nu\nu'} = (-1)^{|i-j|} t_{ij}^{{\bar {\nu'}}{\bar \nu}},
\end{equation}
where $|i-j|$ is the Manhattan distance between site $i$ and $j$.
For these conditions,
the particle-hole transformation
to ensure the positivity of Monte Carlo weights is given by
\begin{eqnarray}
c_{i\nu\uparrow} &\rightarrow& 
(-1)^{i} {\tilde c}_{i{\bar \nu}\uparrow}^{\dagger} \\
c_{i\nu\downarrow} &\rightarrow& 
{\tilde c}_{i\nu\downarrow}.
\end{eqnarray}

\begin{figure}
\hfil
\epsfxsize=8cm
\hfil
\epsfbox{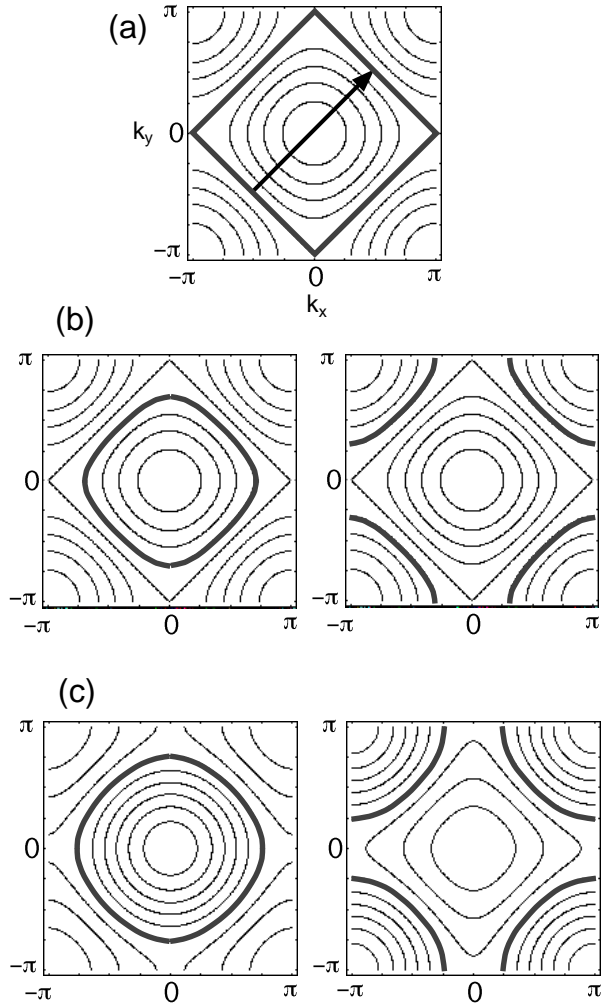}
\caption{
Effects of a particular type of the level split or
long-range hopping at half filling (See text).
(a) Energy contour curves
for the case with only orbitally-diagonal hopping
between nearest neighbor pairs.
The arrow represents the nesting vector $(\pi,\pi)$.
(b) Energy contour curves for each orbital
$\nu=1$(left) and $2$(right)
when the level split $\varepsilon=0.8t$.
(c) Energy contour curves
for each orbital $\nu=1$(left) and $2$(right)
when the next-nearest neighbor hopping $t'=0.2t$.
In all the figures,
the Fermi surfaces are denoted by the bold black curves.
}
\label{etpeff}
\end{figure}

Here, we illustrate these conditions i $\sim$ iv
expressed by Eqs. (\ref{nosign1}) $\sim$ (\ref{nosign4})
and their significance by using a heuristic example.
We consider here a system
with doubly degenerate orbitals in two dimensions.
An effect of finite
$\varepsilon \equiv \varepsilon_{1} = - \varepsilon_{2}$
is easily understood
in the case of non-interacting system
with only the nearest neighbor hoppings
which are diagonal in orbital indices.
Dispersions of two orbitals
are separated by $2\varepsilon$ around the Fermi energy,
therefore, a self doping occurs between two orbitals.
This leads to difference of the Fermi volume and
changes of the nesting behavior
as shown in Fig. \ref{etpeff} (b).
Next, we discuss an effect of long-range hopping.
We consider here the next-nearest neighbor hopping
$t_{\langle\langle ij \rangle\rangle}^{11} =
- t_{\langle\langle ij \rangle\rangle}^{22} = t' \ge 0$,
where $\langle\langle ij \rangle\rangle$ denotes that
the sites $i$ and $j$ are the next-nearest neighbor pairs.
Note that the change in this parameter also
controls the bandwidth.
Fig. \ref{etpeff} (c) shows an effect of
this type of long-range hopping.
In this case also,
a self doping between two orbitals occurs.
In contrast to the case of a finite $\varepsilon$
in Fig. \ref{etpeff} (b),
it should be noted that
long-range hoppings can change the shapes of the Fermi surfaces
as shown in Fig. \ref{etpeff} (c).

It should be stressed here that
these changes by $\varepsilon$ or $t'$ do not break
the perfect nesting property with the nesting vector $(\pi,\pi)$,
because the Fermi surface of the orbital $\nu=1$ coincides with
that of $\nu=2$ by the shift of the momentum $(\pi,\pi)$.
In other words, if the particle-hole symmetry holds,
all the $k$-points on the Fermi surface
are nested with those on the Fermi surface of the other orbital.
This may lead to the Umklapp scattering which may let the system
be insulating when we switch on the interactions.
However, by these changes, the symmetry about the orbital index
which exists at $\varepsilon=t'=0$ is broken and
the self-doping is realized.
It is a non-trivial problem
how the nature of the Mott insulator at $\varepsilon=t'=0$
changes by controlling $\varepsilon$ or $t'$.

Especially, the control of the level split $\varepsilon$ contains
an interesting problem.
The system exhibits two different states as limiting cases;
The Mott insulator at $\varepsilon=0$ and
the band insulator at $\varepsilon=\infty$.
The former is a consequence of the strong correlation, that is,
it does not appear in the non-interacting case.
The latter is a trivial insulator
which has the same energy gap
in both spin and orbital channels as the particle-hole excitation gap.
It deserves studying
how the spin and orbital states change with the level split.
In Sec. \ref{levelcontrol} and \ref{crossover},
we investigate this problem in detail.

\section{Doubly Degenerate Models}
\label{Sec:DDM}

In the following sections,
we concentrate on the doubly degenerate cases of the Hamiltonian
Eq. (\ref{generalH}).
By the auxiliary field QMC technique
introduced in Sec. \ref{Sec:Method},
the ground state properties of doubly degenerate systems are
investigated in one dimension.
In Sec. \ref{DDM},
we define the explicit form of the Hamiltonian
and give some remarks on basic properties of it.
Experimental relevance of our model is discussed in Sec. \ref{ER}.

%%%%%%%%%%
\subsection{Hamiltonian and some remarks}
\label{DDM}

We consider in the following
doubly degenerate Hubbard models,
given by Eq. (\ref{generalH}) with $N_{\rm C} = 2$.
For simplicity, we take into account only nearest neighbor
hoppings which are diagonal about orbital indices.
The level split between orbital $\nu=1$ and $2$ is taken
to satisfy the condition Eq. (\ref{nosign2}).
Moreover, as described in Sec. \ref{Sec:Assumption},
for the convenience of the HS transformation
in the auxiliary field technique,
we take uniform values of $U_{\nu\nu'}$ irrespective of orbital indices.
Then, the explicit forms of Hamiltonian are given by
\begin{eqnarray}
\label{dhmHt}
{\cal H}_{t} &=& -t \sum_{\nu=1}^{2} \sum_{\sigma}
                 \sum_{\langle ij \rangle}
                 c_{i\nu\sigma}^{\dagger} c_{j\nu\sigma} +
                 {\rm h.c.}
\\
\label{dhmHe}
{\cal H}_{\varepsilon} &=& \varepsilon \sum_{i}
                           \left( n_{i1} - n_{i2} \right)
\\
\label{dhmHU}
{\cal H}_{U} &=& U \sum_{i} \sum_{\nu\leq\nu'=1}^{2}
                   \sum_{\sigma\leq\sigma'}
                   \left(1-\delta_{\nu\nu'}\delta_{\sigma\sigma'}\right)
                   n_{i\nu\sigma} n_{i\nu'\sigma'}
\\
\label{dhmHJ}
{\cal H}_{J} &=& 
%                 \frac{J}{2} 
                 J 
                   \sum_{i} \sum_{\nu\neq\nu'=1}^{2}
%                   \sum_{\sigma\sigma'}
                   \sum_{\sigma\leq\sigma'}
                   \bigl( c_{i\nu\sigma}^{\dagger} c_{i\nu'\sigma'}^{\dagger}
                          c_{i\nu\sigma'} c_{i\nu'\sigma}
%\\ \nonumber & & \qquad \qquad \qquad \qquad
                        + c_{i\nu\sigma}^{\dagger} c_{i\nu\sigma'}^{\dagger}
                          c_{i\nu'\sigma'} c_{i\nu'\sigma} \bigr),
\end{eqnarray}
where summations on $\langle ij \rangle$
are over the nearest neighbor sites.
In the following,
only half-filled cases are discussed.
Therefore, the particle-hole symmetry holds
for all the parameter regions,
which ensures the absence of the negative sign difficulty
in the Monte Carlo calculations as explained in Sec. \ref{Sec:Conditions}.

Although this model 
represents a limited case of the general Hamiltonian
(\ref{generalH}),
it still keeps many general aspects and serves our purposes.
The hopping diagonal in orbital indices does not cause
the mixing of two orbitals,
while the last term ${\cal H}_{J}$ contains off-diagonal elements
which mix the orbitals $\nu=1$ and $2$.
%The orbital independence of
%the Coulomb interaction $U$ in Eq. (\ref{dhmHU})
%may not affect the criticality
%because the ${\cal H}_{J}$ term lifts the degeneracy
%which exists at $J=0$.
Note that the second term in Eq. (\ref{dhmHJ})
which denotes the pair hopping term between two orbitals
has been looked over in most of previous works,
although this term is naturally derived from the tight-binding scheme
as mentioned in Sec. \ref{Sec:Introduction}.

Our model Eqs. (\ref{dhmHt}) $\sim$ (\ref{dhmHJ})
is easily shown to have two insulating states as particular limits:
(a) When $U, J \gg t$ at $\varepsilon=0$,
the system is the Mott insulator.
%In this limit, the Hamiltonian is rewritten as
%the $S=1$ spin system by the second order perturbation in $t/U$.
The energy gap, which is called the Mott gap, becomes $2U$.
(b) When $\varepsilon \rightarrow \infty$
with finite $U,J$ and $t$,
since all the electrons fully occupy the orbital $\nu=2$,
the system becomes the band insulator.

For finite values of $U$ and $J$,
the system is expected to be in the Mott insulating state
at $\varepsilon=0$
because of the perfect nesting property.
In the following,
the ground state properties of our model are discussed
by changing the chemical potential $\mu$ and
the level split $\varepsilon$ from this Mott insulator.
By the former control, 
we investigate the critical nature of the Mott transition.
The latter control,
from a consideration in the weak correlation regime
as detailed in Sec. \ref{Sec:Conditions},
leads to the self doping between two orbitals,
which affects spin and/or orbital states in the Mott insulator.
We discuss how the Mott insulator at $\varepsilon=0$
crossovers into the band insulator at $\varepsilon=\infty$
by this self doping in detail.

%%%%%%%%%%
\subsection{Experimental situation}
\label{ER}

In this part, we discuss experimental situations
where our model Eqs. (\ref{dhmHt}) $\sim$ (\ref{dhmHJ}) may have relevance.
As mentioned in the previous section,
the keywords of our model are
two active orbitals and half filling.

In $d$ electron systems, the fivefold energy levels of
the bare $d$ electron is split by the crystal field.
For example, in the octahedral or tetrahedral surroundings,
the fivefold degeneracy splits into the twofold $e_{g}$ and
the threefold $t_{2g}$ orbitals.
For the former octahedral arrangement, the $e_{g}$ levels have
higher energy than the $t_{2g}$ ones,
and for the latter tetrahedral case, upside down.
Therefore, the on-site situation of our model is realized for
the $d^{8}$ configuration in the octahedral surroundings or
the $d^{2}$ configuration in the tetrahedral surroundings.
For both situations,
the level split between the active twofold orbitals
is easily induced by the deformation of the surrounding ligands.
This deformation is considered to be caused
by the lattice structure itself or
the effect of the external pressure.

There are several materials which consist of
the above-mentioned $d^{8}$ units in the octahedral ligand field.
One realization is the group of Ni$^{2+}$ compounds;
KNiF$_{3}$, K$_{2}$NiF$_{4}$,
\cite{Goodenough1963}
Ni(C$_{2}$H$_{8}$N$_{8}$)$_{2}$NO$_{2}$(ClO$_{4}$) (NENP)
\cite{Renard1987,Renard1988}
and its related materials
or Y$_{2}$BaNiO$_{5}$ \cite{Cheong1992,Darriet1993}.
Since in the following sections we investigate our model
in one dimension,
we focus here the last two materials, NENP or Y$_{2}$BaNiO$_{5}$.
These materials have one-dimensional network of the octahedrons
which may be described by our model Hamiltonian
when we forget the ligands.
In these systems, however,
the hopping matrix is not an orbital-diagonal one
like Eq. (\ref{dhmHt});
It may depend on the orbital indices according to each lattice structure.
Moreover, the level split by the deformation of octahedrons
may generally take place in a complicated way, not in a symmetric form
like Eq. (\ref{dhmHe}).
Nevertheless, our simple model Eqs. (\ref{dhmHt}) $\sim$ (\ref{dhmHJ})
serves to investigate the universal aspects of doubly degenerate systems
not depending on details of parameters
as mentioned in the previous section.
In Sec. \ref{Sec:Discussion},
our numerical results will be examined from this point of view
and the relevance on the experimental situations will be discussed.

\section{QMC Results}
\label{Sec:Results}

In this section,
the ground state properties of the doubly degenerate model
defined in Sec. \ref{DDM}
are investigated by the auxiliary field QMC technique
detailed in Sec. \ref{Sec:Method}.
First, numerical details in the QMC calculations are mentioned,
followed by results of this model.

%%%%%%%%%%
\subsection{Details of computations}
\label{NC}

The model defined in Sec. \ref{DDM}
is investigated in one dimension
under the periodic boundary condition.
The QMC data are obtained from $1,000\sim10,000$ sample averages
for the state $\exp(-\frac{\beta}{2}{\cal H})|\phi\rangle$,
where we take $\beta t=10\sim50$ depending on each situation
to obtain converged values in the ground state.
Measurements are divided into five blocks and
the statistical error of a quantity is estimated
by the variance among these five blocks.
In order to accelerate the convergence \cite{Furukawa1991a},
we use the unrestricted Hartree-Fock ground state
as a trial state $|\phi\rangle$.
The QMC calculations are mainly done
for 
$U=4J=3t$.
We set
$\Delta\tau t=0.04$ for
these interaction parameters,
which ensures within the statistical errors
the convergence to the limit $\Delta\tau \rightarrow 0$
for all the physical quantities we calculated.
Detailed discussions of numerical convergence
about $\beta$ and $\Delta\tau$
are given in Appendix A.

%%%%%%%%%%
\subsection{Mott insulating state}
\label{NMIS}

Here we study properties of the Mott insulator at half filling.
Low energy excitations in charge, spin and orbital degrees of freedom
are investigated quantitatively.

First, we calculate the charge gap amplitude
in the Mott insulator at $\varepsilon=0$.
The charge gap is defined as a change of
the total energy $E_{\rm g}$
when we put an extra particle to the ground state of the system
with electron number $N_{\rm e}$,
\begin{equation}
\label{chgapdef0}
\Delta_{\rm C} = E_{\rm g} \left(N_{\rm e}+1\right) -
E_{\rm g} \left(N_{\rm e}\right).
\end{equation}
In actual QMC calculations,
the charge gap is well estimated by the imaginary time dependence of
a single-particle Green function defined as
\begin{equation}
G_{\nu\nu'\sigma} \left(r_{ij},\tau\right) =
\langle \hat{T} c_{i\nu\sigma} \left(\tau\right)
c_{j\nu'\sigma}^{\dagger} \left(0\right) \rangle,
\end{equation}
where $\hat{T}$ gives a time ordered product and
$c_{i\nu\sigma}\left(\tau\right)= e^{\tau{\cal H}}
c_{i\nu\sigma} e^{-\tau{\cal H}}$
is in the Heisenberg representation.
The bracket defines the canonical ensemble average.
The Green function in the large $\tau$ limit is governed by the charge gap,
which is explicitly shown
by the spectral representation as
\begin{eqnarray}
& & \lim_{\tau\rightarrow\infty} G_{\nu\nu'\sigma} \left(r_{ij},\tau\right)
\nonumber \\
&=&
\lim_{\tau\rightarrow\infty} \sum_{m}
\langle \Psi_{\rm g}^{N_{\rm e}} | c_{i\nu\sigma}
|  \Psi_{m}^{N_{\rm e}+1} \rangle
%\\ \nonumber & & \qquad \qquad \quad
\langle \Psi_{m}^{N_{\rm e}+1} | c_{j\nu'\sigma}^{\dagger}
|  \Psi_{\rm g}^{N_{\rm e}} \rangle
%\\
%& & \qquad \qquad
% \qquad
\exp\left[ -\tau \{ E_{m}\left(N_{\rm e}+1\right)
                      - E_{\rm g}\left(N_{\rm e}\right) \} \right]
\nonumber \\
&\sim&
\langle \Psi_{\rm g}^{N_{\rm e}} | c_{i\nu\sigma}
|  \Psi_{\rm g}^{N_{\rm e}+1} \rangle
%\\ \nonumber & & \qquad
\langle \Psi_{\rm g}^{N_{\rm e}+1} | c_{j\nu'\sigma}^{\dagger}
|  \Psi_{\rm g}^{N_{\rm e}} \rangle
\label{Gexpdecay}
%\\
%& & \qquad \qquad
% \qquad
\exp\left[ -\tau \{ E_{\rm g}\left(N_{\rm e}+1\right)
                      - E_{\rm g}\left(N_{\rm e}\right) \} \right],
\nonumber
\end{eqnarray}
where $|\Psi_{m}^{N_{\rm e}}\rangle$
and $E_{m}\left(N_{\rm e}\right)$
represent eigenvectors and eigenvalues in the $N_{\rm e}$ electron system
respectively and $|\Psi_{\rm g}\rangle$ is the ground state.
We use the numerically stable algorithm
recently proposed by Assaad and Imada
to calculate the imaginary time dependence of the Green functions
\cite{Assaad1996b}.

\begin{figure}
\hfil
\epsfxsize=8cm
\hfil
\epsfbox{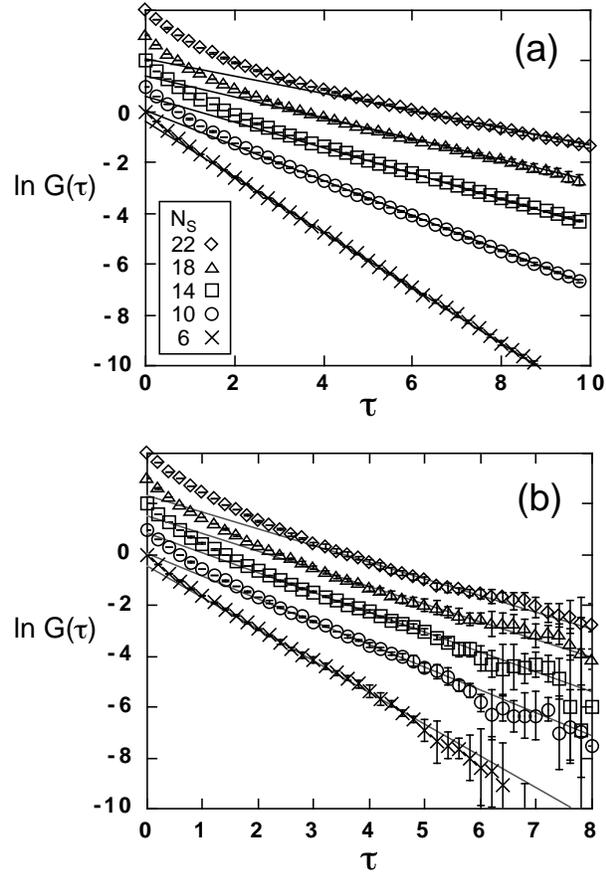}
\caption{
Imaginary time dependence of Green functions at $\varepsilon=0$;
(a) for $U=4J=2t$ and (b) for $U=4J=3t$.
Note that the origin for the $y$ axis has an offset
to distinguish the data
for each system size in both figures.
The straight lines are the fit for each data.
}
\label{FIGGtau}
\end{figure}

\begin{figure}
\hfil
\epsfxsize=8cm
\hfil
\epsfbox{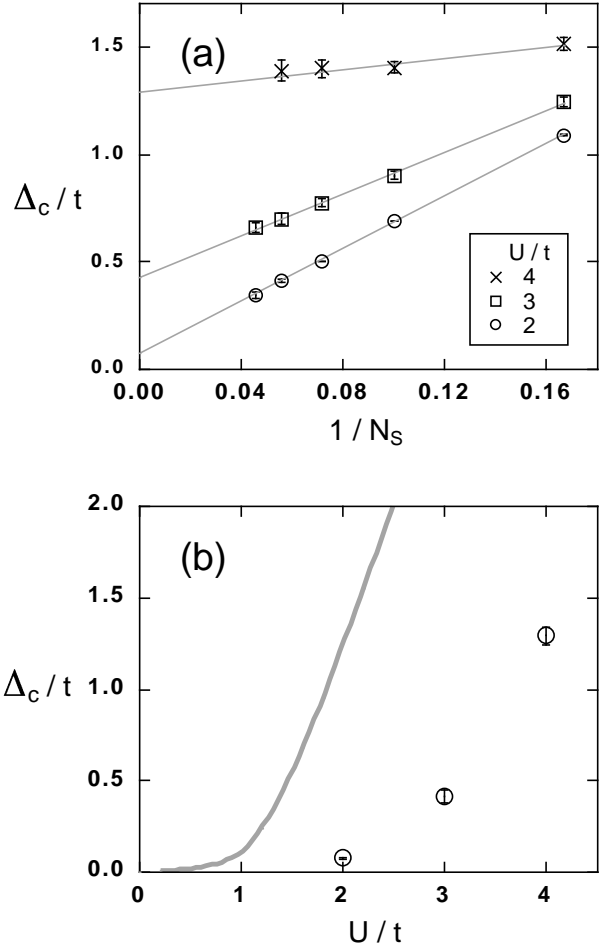}
\caption{
QMC data for the charge gap.
We fix $U/J=4$.
(a) Extrapolations of the charge gap to the infinite system size.
(b) The charge gap in the thermodynamic limit
for various values of $U/t$ at $\varepsilon=0$.
The gray curve is the result of the Hartree-Fock approximation.
}
\label{FIGgapExt}
\end{figure}

Fig. \ref{FIGGtau} shows typical imaginary-time dependence
of a uniform Green function defined as
\begin{equation}
\label{uniformG}
G\left(\tau\right) = \frac{1}{N_{\rm S}}
\sum_{i\nu\sigma} G_{\nu\nu\sigma}\left(r_{ii},\tau\right).
\end{equation}
The charge gap amplitude for each system size is determined by
the least square fit for
the exponential tail of $G\left(\tau \right)$.
The fitting lines are shown in the figures as the straight lines.
The values of $\Delta_{\rm C}$ in the thermodynamic limit
are obtained by an extrapolation in the inverse of system sizes
as shown in Fig. \ref{FIGgapExt} (a).
The QMC data are well fitted by the linear function of $1/N_{\rm S}$.
The reason for these linear relations is not apparent as it stands.
However, this may have a connection with
a gapless nature in orbital degrees of freedom,
which is mentioned below.

At $\varepsilon=0$, the charge gap as function of interaction strength
$U/t$ is summarized in Fig. \ref{FIGgapExt} (b).
For the comparison, the Hartree-Fock result
is also plotted in the figure.
Our QMC results show a large reduction from the mean-field results.
Our model has both spin and orbital components and
there is some degeneracy about order parameters.
The simple mean-field approach may become crude for this type of systems
because it tends to favor one ordering among degenerate order parameters
to open the energy gap and
neglect the competition among them. 
See Appendix C for details of the mean-field results.

%\begin{table}
%\caption{QMC results of the charge gap in the thermodynamic limit
%for various values of $U/t$ at $\varepsilon=0$.
%We fix $U/J=4$.
%Errorbars in the last digits are given in the parenthesis.}
%\begin{center}
%\begin{tabular}{@{\hspace{\tabcolsep}\extracolsep{\fill}} cr} \hline
%$U/t$ & $\Delta_{\rm C}/t$ \\ \hline
%2 & 0.075(3) \\
%3 & 0.42(2) \\
%4 & 1.29(5) \\ \hline
%\end{tabular}
%\end{center}
%\end{table}

\begin{figure}
\hfil
\epsfxsize=8cm
\hfil
\epsfbox{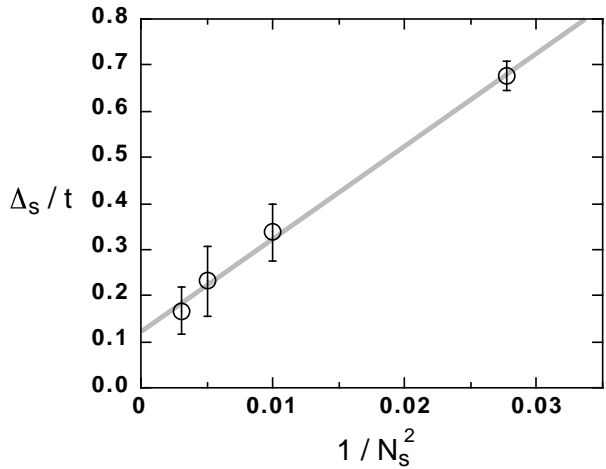}
\caption{
QMC data for the size dependence of
the spin gap at $U=4J=4t$ and $\varepsilon=0$.
}
\label{FIG:Sgap}
\end{figure}

Next, we discuss spin degrees of freedom.
In Fig. \ref{FIG:Sgap},
we show the QMC results of the spin gap $\Delta_{\rm S}$
calculated as the energy difference
between the singlet ground state and the first triplet excitation,
%\begin{eqnarray}
\begin{equation}
\label{Sgap}
\Delta_{\rm S} 
%&=&
=
 E_{\rm g}
\left(N_{\rm e}^{\uparrow} = \frac{N_{\rm e}}{2} + 1;
      N_{\rm e}^{\downarrow} = \frac{N_{\rm e}}{2} - 1\right)
%\\ \nonumber
%&-&
-
 E_{\rm g}
\left(N_{\rm e}^{\uparrow} = \frac{N_{\rm e}}{2};
      N_{\rm e}^{\downarrow} = \frac{N_{\rm e}}{2}\right).
%\end{eqnarray}
\end{equation}
In the figure, we fit the data by
$\Delta_{\rm S} = {\rm const.} + a N_{\rm S}^{-2}$
which was used to estimate the spin gap
in the $S=1$ spin system with nearest-neighbor antiferromagnetic
coupling, which is often called the Haldane system
\cite{White1992}.
Our data suggest a non-zero value of the spin gap
in the thermodynamic limit;
$\Delta_{\rm S}/t = 0.12 \pm 0.05$.
However, our data are not accurate
enough to justify this form of the fit,
because of large errorbars and
finite size limitation.
Further investigations are necessary to estimate the spin gap
$\Delta_{\rm S}$ accurately.

It should be stressed that
the magnitude of the spin gap $\Delta_{\rm S}$ obtained here
is much smaller than the charge gap $\Delta_{\rm C}$
plotted in Fig. \ref{FIGgapExt} (b);
$\Delta_{\rm S} / \Delta_{\rm C} \sim 0.1$ for $U=4J=4t$.
This large discrepancy between  $\Delta_{\rm C}$ and $\Delta_{\rm S}$
is characteristic for the strongly correlated insulator.
Note that $\Delta_{\rm S}$ is equal to $\Delta_{\rm C}$
in the band insulator.

If we take the different values between intra- and inter-orbital
Coulomb interactions, that is,
$U_{11}=U_{22}>U_{12}$,
an effective Hamiltonian
at $\varepsilon=0$ in the limit of strong correlation
is the Haldane system.
As widely known,
this $S=1$ spin system has been predicted
to have a finite excitation gap
between the singlet ground state and the triplet excited state,
which is called the Haldane gap
\cite{Haldane1983a,Haldane1983b}.
This conjecture has been favorably supported
by both analytical
\cite{Affleck1987,Affleck1988} and numerical
\cite{Nightingale1986,Nomura1989,Kennedy1990,White1992,White1993}
investigations.
Our numerical result obtained above suggests that
the Mott insulating state of our model at $\varepsilon=0$
may be smoothly connected with the Haldane system,
that is, it may be in the same universality class
in spite of finite values of $U$ and $J$,
although our model with the assumption $U_{11}=U_{22}=U_{12}=U$
has an additional degeneracy in the limit of strong correlation.

\begin{figure}
\hfil
\epsfxsize=8cm
\hfil
\epsfbox{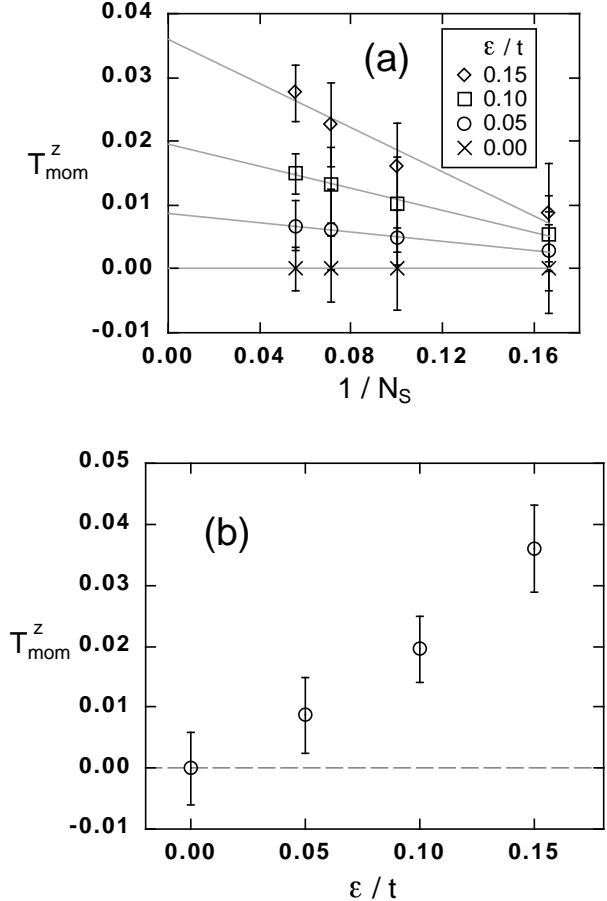}
\caption{
QMC data of the pseudo-spin moment for $U=4J=3t$.
(b) shows the values of the pseudo-spin moment
as function of $\varepsilon$,
which are obtained from 
the extrapolations to the thermodynamic limit in (a).
}
\label{FIGTzmomins}
\end{figure}

Finally, we consider properties in orbital degrees of freedom.
In the doubly degenerate system,
it is useful to describe the two orbital degrees of freedom 
by the pseudo-spin operator defined as
\begin{equation}
\label{ispop}
T_{i}^{\alpha} = \frac{1}{2}
\sum_{\sigma} \sum_{\nu\nu'}
\vec{\tau}_{\nu\nu'}^{\alpha}
c_{i\nu\sigma}^{\dagger} c_{i\nu'\sigma},
\end{equation}
where $\vec{\tau}$ is the Pauli matrix and $\alpha = x$, $y$ or $z$.
In order to discuss the low-lying excitations in this channel,
we calculate how the pseudo-spin moment
responds to its conjugate field, that is, the level split $\varepsilon$.
The pseudo-spin moment is defined as
\begin{equation}
\label{Tzmom}
T_{\rm mom}^{z} = \frac{1}{N_{\rm S}} | \sum_{i} T_{i}^{z} |,
\end{equation}
which signals a magnitude of self doping between two orbitals.
As shown in Fig. \ref{FIGTzmomins} (a),
the pseudo-spin moment $T_{\rm mom}^{z}$
has a finite value in the thermodynamic limit
when we switch on $\varepsilon$.
Fig. \ref{FIGTzmomins} (b) summarizes the pseudo-spin moment $T_{\rm mom}^{z}$
as function of $\varepsilon$.
We find that
$T_{\rm mom}^{z}$ seems to grow continuously
from zero at $\varepsilon=0$,
which strongly suggests a gapless nature in the orbital channel
in the Mott insulating state.

%%%%%%%%%%
\subsection{Mott transition by chemical potential control}
\label{MTCPC}

We consider here insulator-metal transitions
by controlling the chemical potential $\mu$.
Critical exponents are investigated from the insulating side
by a technique recently proposed and applied to
single-orbital Hubbard models \cite{Assaad1996a}.
In this method,
the correlation length exponent $\nu$ is directly determined
from the localization length $\xi_{\rm l}$ of
the single-particle wave function
probed by of the Green function in the long distance.
Here, we explain the prescription to determine the exponent $\nu$.

\begin{figure}
\hfil
\epsfxsize=8cm
\hfil
\epsfbox{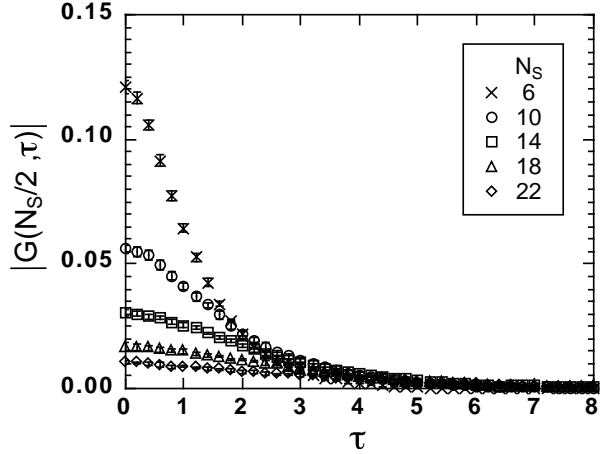}
\caption{
Imaginary time dependence of Green functions for the farthest sites
at $U=4J=3t$ and $\varepsilon=0$.
}
\label{FIGGR}
\end{figure}

Fig. \ref{FIGGR} shows the imaginary time dependence
of the Green function for the farthest sites
$G(r_{ij}=N_{\rm S}/2,\tau) \equiv \sum_{\nu\sigma}
G_{\nu\nu\sigma}(r_{ij}=N_{\rm S}/2,\tau)$
for each system size $N_{\rm S}$ at $\varepsilon=0$.
From these data, we obtain the frequency dependent Green function
for $|\mu|<\Delta_{\rm C}$ as
\begin{equation}
\label{FTG}
G\left(r_{ij},\omega=\mu\right) =
\int_{-\infty}^{\infty} d\tau
G\left(r_{ij},\tau\right) e^{\tau\mu},
\end{equation}
because the density is fixed in the Mott insulating state.
It should be noted that
the Green function satisfies
$G(r_{ij},\tau) = (-1)^{|r_{ij}|} G(r_{ij},-\tau)$
by the particle-hole symmetry.
In Eq. (\ref{FTG}),
the factor $e^{\tau\omega}$ makes it difficult to estimate
$G\left(r_{ij},\omega=\mu\right)$ accurately,
because the statistical error with each QMC data will grow
exponentially with increasing values of $\tau$ for $\tau\omega>0$.
For better numerical estimation,
we carry out the imaginary time integration in Eq. (\ref{FTG})
in the following way \cite{AssaadPREP}.
Using the operator $T_{R}$ which satisfies
$T_{R}^{\dagger} c_{i\nu\sigma} T_{R} = c_{i+R,\nu\sigma}$,
we describe the large $\tau$ behavior of
$G\left(N_{\rm S}/2,\tau\right)$ as
%\begin{eqnarray}
\begin{equation}
%& &
G\left(N_{\rm S}/2,\tau\right)
%\nonumber \\
%&\sim&
\sim
 \sum_{\nu\sigma}
\langle \Psi_{\rm g}^{N_{\rm e}} |
T_{N_{\rm S}/2}^{\dagger} c_{i\nu\sigma} T_{N_{\rm S}/2}
|  \Psi_{\rm g}^{N_{\rm e}+1} \rangle
%\\ \nonumber & & \qquad
\langle \Psi_{\rm g}^{N_{\rm e}+1} | c_{i\nu\sigma}^{\dagger}
|  \Psi_{\rm g}^{N_{\rm e}} \rangle
\exp\left( -\tau\Delta_{\rm C} \right)
%\end{eqnarray}
\end{equation}
in the spectral representation as Eq. (\ref{Gexpdecay}).
Here, our Hamiltonian commutes with the operator $T_{N_{\rm S}/2}$ and
we employ the periodic boundary conditions
$T_{N_{\rm S}/2}^{2} = 1$.
Therefore, $G\left(N_{\rm S}/2,\tau\right) \sim
\pm G\left(\tau\right)$ for large $\tau$,
because
$T_{N_{\rm S}/2} | \Psi_{\rm g}^{N_{\rm e}} \rangle
= \pm | \Psi_{\rm g}^{N_{\rm e}} \rangle$.
Here $G(\tau)$ is the uniform Green function
defined in Eq. (\ref{uniformG}).
With this property, we estimate
the integration in Eq. (\ref{FTG}) by substituting
the large $\tau$ behavior of $G\left(\tau\right)$,
that is, $\alpha  \exp\left( -\tau\Delta_{\rm C} \right)$, 
for that of $G\left(N_{\rm S}/2,\tau\right)$, as
%\begin{eqnarray}
\begin{equation}
%& &
 |G\left(r_{ij},\omega=\mu\right)|
%\nonumber \\
%&=&
=
\int_{-\tau_{0}}^{+\tau_{0}} d\tau
|G\left(N_{\rm S}/2,\tau\right)| e^{\tau\mu}
%\nonumber \\
%&+&
+
 \alpha \left[
\int_{+\tau_{0}}^{+\infty} e^{\tau\left(\mu-\Delta_{\rm C}\right)}
+
\int_{-\infty}^{-\tau_{0}} e^{\tau\left(\mu+\Delta_{\rm C}\right)}
\right].
\label{GRfit}
%\end{eqnarray}
\end{equation}
Here, we take the threshold value $\tau_{0}$ such that
for $\tau>\tau_{0}$ the tail of $G\left(N_{\rm S}/2,\tau\right)$
obeys $\alpha \exp\left( -\tau\Delta_{\rm C} \right)$
within our numerical accuracy.
The advantage in Eq. (\ref{GRfit}) lies in the fact that
$G\left(\tau\right)$ generally shows better numerical convergence
for large $\tau$ than
$G\left(N_{\rm S}/2,\tau\right)$.

\begin{figure}
\hfil
\epsfxsize=8cm
\hfil
\epsfbox{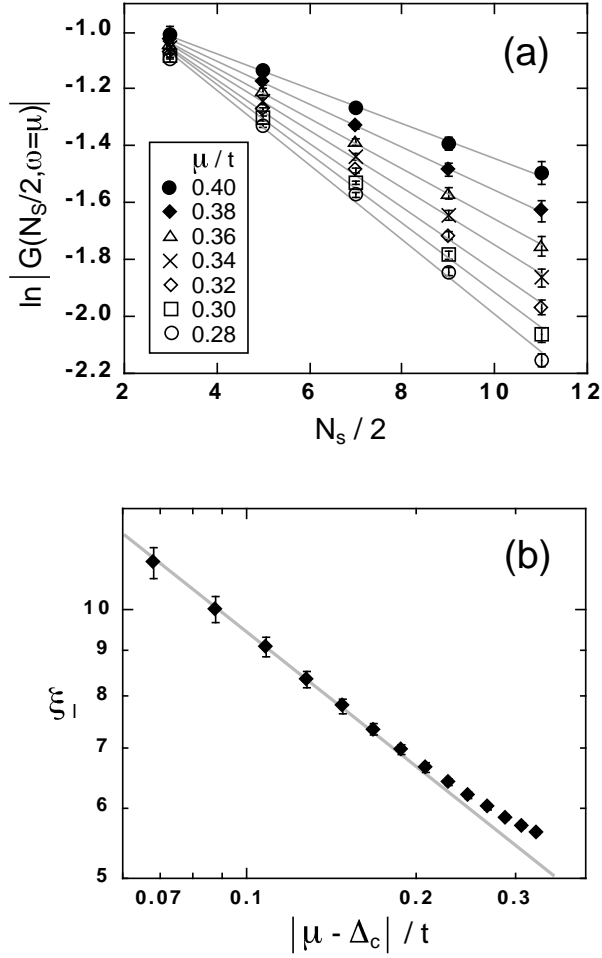}
\caption{
Schemes to calculate the localization length $\xi_{\rm l}$
and the correlation length exponent $\nu$. See text for details.
(a) Distance dependence of the Green function
for various values of the chemical potential $\mu$.
(b) Algebraic divergence of the localization length
toward the critical point.
The gray line corresponds to 
$\xi_{\rm l} \sim |\mu-\Delta_{\rm C}|^{-1/2}$.
}
\label{FIGgzil}
\end{figure}

Fig. \ref{FIGgzil} (a) represents 
$N_{\rm S}$ dependence of $G(N_{\rm S}/2,\omega=\mu)$
at $\varepsilon=0$ for various values of the chemical potential $\mu$
obtained by the above prescription.
For each value of $\mu$,
$G\left(N_{\rm S}/2,\mu\right)$ decays exponentially
with $N_{\rm S}$ due to a finite localization length
$\xi_{\rm l}$ in the insulating state, that is,
$G(r_{ij},\mu) \sim \exp(-|r_{ij}|/\xi_{\rm l})$.
We determine the localization length $\xi_{\rm l}$
from the decay rates in Fig. \ref{FIGgzil} (a).
Fig. \ref{FIGgzil} (b) gives a log-log plot between $\xi_{\rm l}$ and
$|\mu-\Delta_{\rm C}|$,
which determines the critical exponent $\nu$ directly
through a scaling relation
\begin{equation}
\label{sr}
\xi_{\rm l} \sim |\mu-\Delta_{\rm C}|^{-\nu}.
\end{equation}
We note that the charge gap is determined from Eq. (\ref{chgapdef0})
independently of this procedure.
%This relation will be discussed again in Sec. \ref{dhmdisc:scaling}.
In this case with $\varepsilon=0$,
the QMC data may be consistently understood by $\nu=1/2$.
We will discuss this result as well as
the scaling relation Eq. (\ref{sr})
in Sec. \ref{dhmdisc:scaling}.

%%%%%%%%%%
\subsection{Self doping by level split control}
\label{levelcontrol}

In this section, we concentrate on the control of
the level split $\varepsilon$.
When we change the level split $\varepsilon$,
the self doping between two orbitals occurs continuously
from $\varepsilon=0$,
because of the gapless nature in pseudo-spin degrees of freedom
discussed in Sec. \ref{NMIS}.

\begin{figure}
\hfil
\epsfxsize=8cm
\hfil
\epsfbox{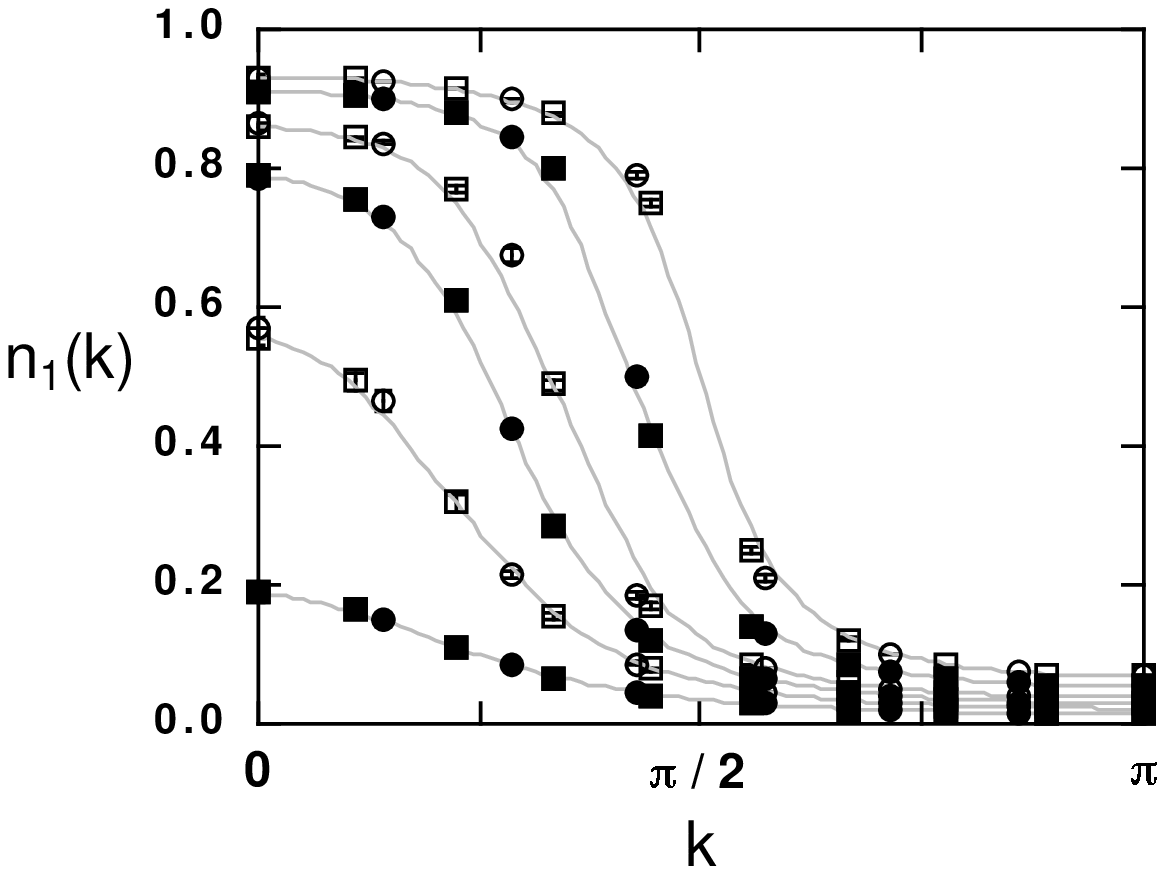}
\caption{
Momentum distribution functions of the orbital $\nu=1$
for the level split control at $U=4J=3t$.
The gray curves are guides to eye.
The data are for $\varepsilon/t=0.0, 0.4, 0.8, 1.0, 1.2$ and $1.4$
from the top to the bottom.
The circles and squares correspond to the data for
$N_{\rm S} = 14$ and $18$, respectively.
}
\label{FIGnk}
\end{figure}

Fig. \ref{FIGnk} shows the typical behaviors of
the momentum distribution function defined as
\begin{equation}
\label{nkdef}
n_{\nu} \left(k\right) = \frac{1}{N_{\rm S}}
\sum_{ij\sigma} \langle c_{i\nu\sigma}^{\dagger}
c_{j\nu\sigma} \rangle e^{{\rm i}kr_{ij}}.
\end{equation}
At $\varepsilon=0$, $n_{\nu=1}(k)$ is equal to $n_{\nu=2}(k)$
because of the symmetry about the orbital index,
and the characteristic wave number $k^{*}$
where $n_{1}(k)$ has the largest slope is at $\pi/2$.
When $\varepsilon$ becomes larger,
the form of $n_{1}(k)$ changes continuously
and the wave number $k^{*}$ approaches zero.
Note that the relation $n_{2}(k) = 1- n_{1}(\pi-k)$ holds
for all the values of $\varepsilon$
by the particle-hole symmetry.
The area under the curve of $n_{\nu}(k)$ corresponds to
the density in the orbital $\nu$,
therefore these behaviors of the momentum distribution function
clearly represent the self doping
from the orbital $\nu=1$ to $\nu=2$.

\begin{figure}
\hfil
\epsfxsize=8cm
\hfil
\epsfbox{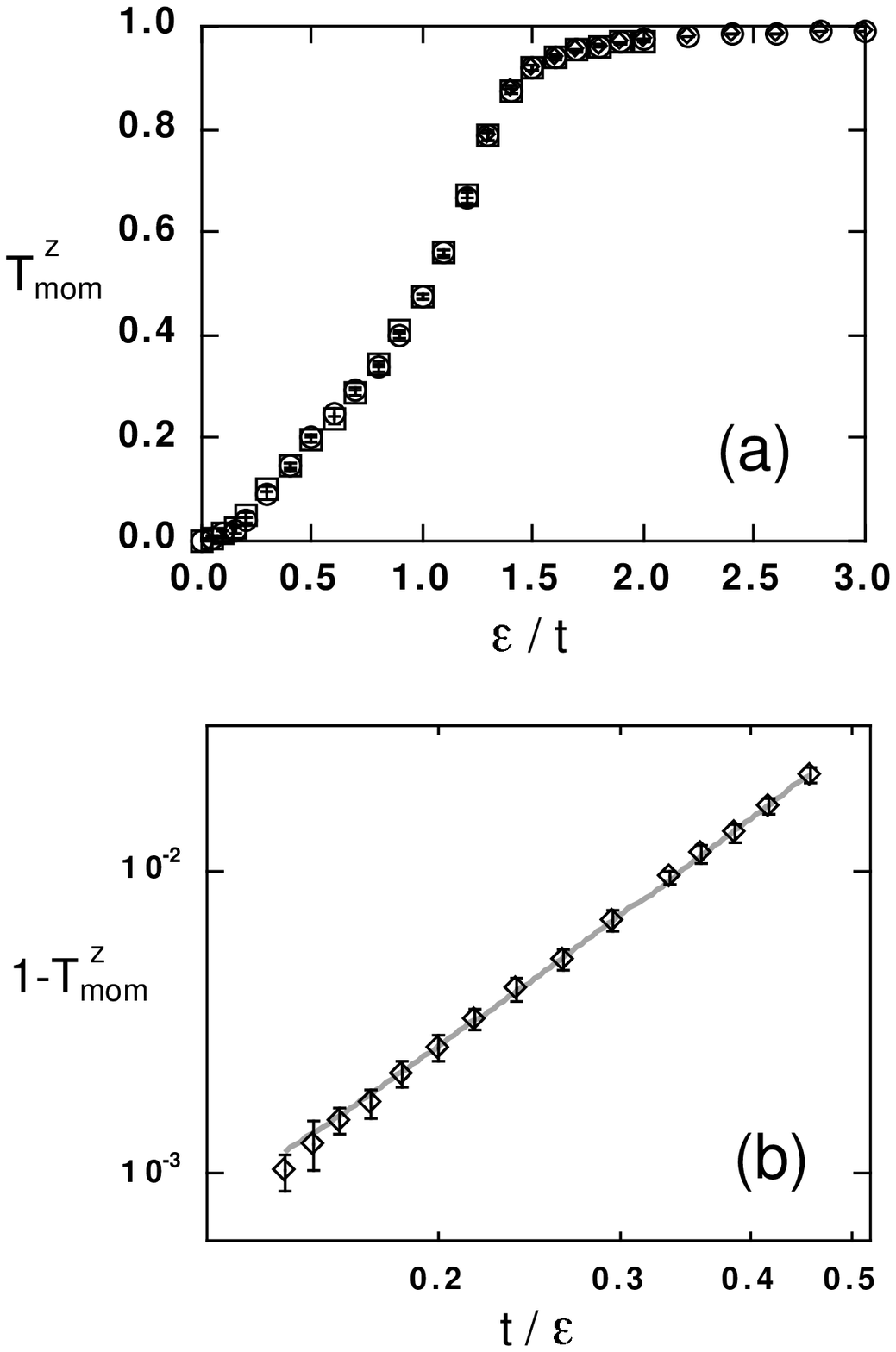}
\caption{
Pseudo-spin moment
for the level split control at $U=4J=3t$.
The asymptotic behavior of the self doping for large values of
$\varepsilon$ is illustrated in (b).
The gray curve is the fit by
$a/\varepsilon^{2} + b/\varepsilon^{3}$.
In both figures,
the diamonds, circles and squares correspond to the data for
$N_{\rm S} = 10$, $14$ and $18$, respectively.
}
\label{FIGTzmom}
\end{figure}

Fig. \ref{FIGTzmom} shows
the $\varepsilon$ dependence of the pseudo-spin moment
defined in Eq. (\ref{Tzmom}).
As shown in Fig. \ref{FIGTzmom} (a),
the pseudo-spin moment $T_{\rm mom}^{z}$ grows continuously
from zero at $\varepsilon=0$ and
approaches $1$ for large values of $\varepsilon$.
The behavior of $T_{\rm mom}^{z}$ for large $\varepsilon$
is also plotted in Fig. \ref{FIGTzmom} (b)
as function of $\varepsilon^{-1}$.
This asymptotic behavior of
the magnitude of the self doping is discussed in Sec. \ref{crossover}.

\begin{figure}
\hfil
\epsfxsize=8cm
\hfil
\epsfbox{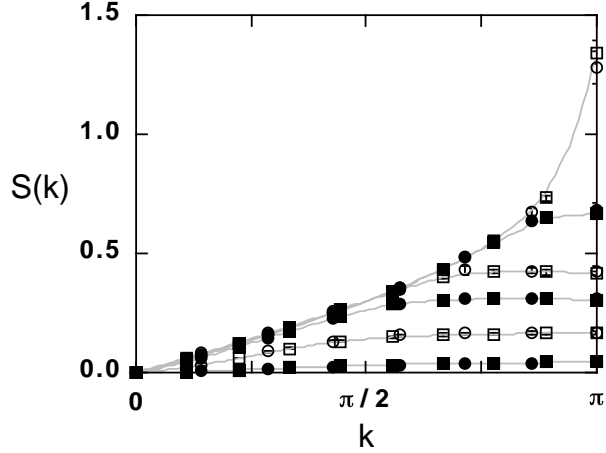}
\caption{
Spin correlation functions
for the level split control at $U=4J=3t$.
The gray curves are guides to eye.
The data are for $\varepsilon/t=0.0, 0.4, 0.8, 1.0, 1.2$ and $1.4$
from the top to the bottom.
The circles and squares correspond to the data for
$N_{\rm S} = 14$ and $18$, respectively.
}
\label{FIGSk}
\end{figure}

\begin{figure}
\hfil
\epsfxsize=8cm
\hfil
\epsfbox{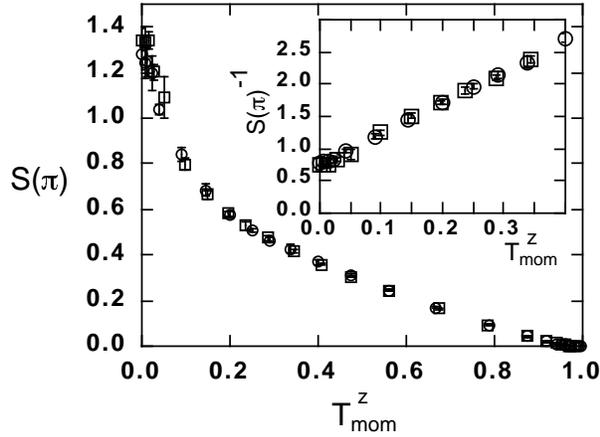}
\caption{
The antiferromagnetic spin correlation as function of
the pseudo-spin moment at $U=4J=3t$.
The inset shows the behavior of the inverse of $S(\pi)$
near $T_{\rm mom}^{z} = 0$.
The circles and squares correspond to the data for
$N_{\rm S} = 14$ and $18$, respectively.
}
\label{FIGSpivssd}
\end{figure}

\begin{figure}
\hfil
\epsfxsize=8cm
\hfil
\epsfbox{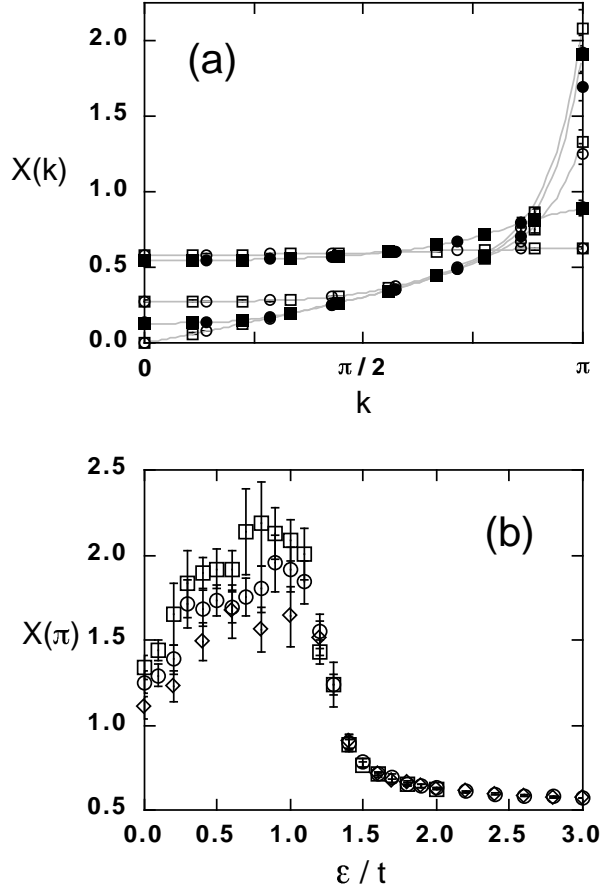}
\caption{
(a) Anomalous correlation functions
for the level split control at $U=4J=3t$.
The gray curves are guides to eye.
The data are for $\varepsilon/t=0.0, 0.6, 1.0, 1.4$ and $2.0$
from the bottom to the top at $k=0$.
(b) $\varepsilon$ dependence of $X(\pi)$.
In both figures,
the diamonds, circles and squares correspond to the data for
$N_{\rm S} = 10, 14$ and $18$, respectively.
}
\label{FIGXk}
\end{figure}

The spin correlation functions
for various values of $\varepsilon$ are shown in Fig. \ref{FIGSk}.
The definition of the spin correlation function is
\begin{equation}
\label{Skdef}
S \left(k\right) = \frac{1}{N_{\rm S}}
\sum_{ij} \langle S_{i} S_{j} \rangle e^{{\rm i}kr_{ij}},
\end{equation}
where $S_{i} = \left(S_{i}^{x} +S_{i}^{y} +S_{i}^{z} \right)/3$.
The spin operator is defined as
\begin{equation}
\label{spop}
S_{i}^{\alpha} = \frac{1}{2}
\sum_{\nu} \sum_{\sigma\sigma'}
\vec{\tau}_{\sigma\sigma'}^{\alpha}
c_{i\nu\sigma}^{\dagger} c_{i\nu\sigma'}
\end{equation}
with the Pauli matrix $\vec{\tau}$ and $\alpha = x$, $y$ or $z$.
At $\varepsilon=0$,
although it has a spin excitation gap
%due to the $S=1$ excitation spectrum of the Haldane system,
as examined in Sec. \ref{NMIS},
$S(k)$ has a large peak at $k=\pi$
originating from the short-range antiferromagnetic correlation
in the Mott insulator.
When we change the level split $\varepsilon$,
this structure collapses and the peak is broadened continuously.

In Fig. \ref{FIGSpivssd},
we summarize the values of the antiferromagnetic correlation
$S(\pi)$ as function of
the pseudo-spin moment $T_{\rm mom}^{z}$
which corresponds to the magnitude of self doping.
The inset of Fig. \ref{FIGSpivssd} plots
the inverse of $S(\pi)$ in the small self-doped region.
These behaviors of the spin correlation for the self doping
are discussed in Sec. \ref{crossover}.

As mentioned in Sec. \ref{Sec:Conditions},
for a finite value of $\varepsilon$,
the nesting property between different orbitals
may play a significant role.
Actually, in the mean-field analysis which is discussed in Appendix C,
the system keeps the insulating nature
when we change $\varepsilon$ from zero, where
the energy gap opens by the order parameter which
corresponds to the scattering process between different spins and orbitals.
This relevant process is represented by the operator
\begin{equation}
X_{i} = \frac{1}{2} 
\sum_{\nu\nu'} \sum_{\sigma\sigma'}
\vec{\tau}_{\nu\nu'}^{x}
\vec{\tau}_{\sigma\sigma'}^{x}
c_{i\nu\sigma}^{\dagger} c_{i\nu'\sigma'}.
\end{equation}
The correlation function of this operator, $X(k)$, is defined as
\begin{equation}
\label{Xkdef}
X \left(k\right) = \frac{1}{N_{\rm S}}
\sum_{ij} \langle X_{i} X_{j} \rangle e^{{\rm i}kr_{ij}}.
\end{equation}
We call this the anomalous correlation function.

In Fig. \ref{FIGXk} (a), we show the behaviors of
the anomalous correlation function $X(k)$
with the change of the level split $\varepsilon$.
Note that in the non-interacting case,
$X(k)$ takes the constant value $1/2$ by definition.
We plot the values of $X(\pi)$ in Fig. \ref{FIGXk} (b).
Compared with the monotonic decrease of the spin correlation $S(\pi)$,
$X(\pi)$ is enhanced for small values of $\varepsilon$ and
keeps almost constant values up to $\varepsilon \sim t$.
This behavior is discussed in Sec. \ref{crossover}
in comparison with the mean-field results obtained in Appendix C.

\section{Discussion}
\label{Sec:Discussion}

In this section, we discuss the numerical results
in one dimension obtained in the previous section.
The Mott transition by the chemical potential control
is analyzed based on the scaling theory.
The behaviors by changing the level split are discussed
as the crossover between the Mott insulator with high spin state
and the band insulator with low spin state.
The relevance of our results to experimental situations
is also remarked.
We also compare our results with
previous theoretical investigations.

%%%%%%%%%%
\subsection{Hyperscaling analysis}
\label{dhmdisc:scaling}

In this part, the QMC data
for the insulator-metal transitions
controlled by the chemical potential
obtained in Sec. \ref{MTCPC} are analyzed
based on the hyperscaling hypothesis.
The universality class is examined
by combining scaling relations with the QMC results.

Note that the hyperscaling analysis assumes the continuous transition.
All the numerical data in our investigations
do not show any apparent discontinuity
expected in first-order transitions.
This implies that the transitions are always continuous here,
although we can not exclude the possibility
of a weak first-order transition with small discontinuity.
In this section, we assume the continuous transitions
and employ the hyperscaling hypothesis.

First, we briefly review scaling arguments
for continuous metal-insulator transitions
%\cite{Continentino1992,Stafford1993,Continentino1994,Imada1995a,Imada1995b}.
\cite{Stafford1993,Imada1995a,Imada1995b}.
The hyperscaling hypothesis asserts that
a single characteristic length scale $\xi$ and
a single energy scale $\Omega$ govern the quantum critical phenomena.
The scaling behaviors of these two quantities give
two basic critical exponents;
the correlation length exponent $\nu$ and
the dynamical exponent $z$, as
\begin{eqnarray}
\label{xiscaling}
\xi &\sim& |\Delta|^{-\nu} \\
\label{Omegascaling}
\Omega &\sim& \xi^{-z} \sim |\Delta|^{z\nu},
\end{eqnarray}
where $\Delta$ is a control parameter for the transition
which measures a distance from the critical point.
With these two exponents $\nu$ and $z$,
the finite-size scaling for the singular part of the free energy density
in $d$ dimensions is given by
\begin{equation}
\label{freesing}
f_{\rm S} \left(\Delta \right) \sim
\Delta^{\nu\left(d+z\right)} F\left( \xi/L, \xi^{z}/\beta \right),
\end{equation}
where $F$ is a finite-size scaling function,
$L$ is a linear dimension of the system and
$\beta$ is the inverse temperature.
Comparing differentiations of $f_{\rm S}$ by $L$ or $\beta$
with the Taylor expansion of the free energy density
for twists of spatial or temporal boundary conditions
with dimensional analysis,
we obtain the scaling behaviors
for the singular part of physical quantities; for example,
\begin{eqnarray}
%\label{Dscaling}
%\mbox{Drude weight} \; &:& \quad
%D \sim \Delta^{\nu\left(d+z-2\right)} \\
\label{csuscep}
\mbox{charge susceptibility} \; &:& \quad
\kappa \sim \Delta^{-\nu\left(z-d\right)} \\
%\label{osuscep}
%\mbox{orbital susceptibility} \; &:& \quad
%\kappa_{\rm O} 
%%\equiv -\frac{\partial^{2}f_{\rm S}}
%%{\partial\varepsilon^{2}}
%\sim \Delta^{-\nu\left(z-d\right)}\\
\label{gzilscaling}
\mbox{localization length} \; &:& \quad
\xi_{\rm l} \sim \Delta^{-\nu}.
%\label{chgapscaling}
%\mbox{charge gap} \; &:& \quad
%\Delta_{\rm C} \sim \Delta^{z\nu}.
\end{eqnarray}
The last one Eq. (\ref{gzilscaling}) is the same as Eq. (\ref{sr}).

It is important to notice that
the charge susceptibility $\kappa$
can also be deduced directly from the definition
$-\partial^{2}f/\partial\Delta^{2}$
where $\Delta$ expresses the chemical potential $\mu$.
Direct differentiation of Eq. (\ref{freesing}) gives
\begin{equation}
\label{suscep2}
\kappa \sim \Delta^{\nu\left(d+z\right)-2}.
\end{equation}
If we compare this relation with Eq. (\ref{csuscep}),
we obtain an important relation
\begin{equation}
\label{zv=1}
z\nu = 1
\end{equation}
for metal-insulator transitions by the chemical potential.

In Sec. \ref{MTCPC},
we studied the correlation length exponent $\nu$ directly
from the insulating side.
The QMC data of the localization length $\xi_{\rm l}$
in Fig. \ref{FIGgzil} (b) suggests $\nu=1/2$.
%through the scaling relation Eq. (\ref{gzilscaling}).
%Therefore, with the general relation for this case Eq. (\ref{zv=1}),
%the Mott transition by the chemical potential control in our model
%is suggested to be in the universality class
%characterized by
%\begin{equation}
%\label{cemyu}
%z=1/\nu=2.
%\end{equation}

The criticality of the filling-control Mott transitions
itself in one dimension
has been investigated based on the scaling theory \cite{Imada1995a}.
There, the critical nature has been discussed depending on
whether component (spin or orbital) orderings exist or not
both in insulators and in metals.
The prediction of these scaling arguments is that
all the Mott transitions by controlling the chemical potential
in one dimension should be
characterized by $\nu=1/z=1/2$.
Our result, $\nu=1/2$ for the doubly degenerate Hubbard model
also support at least this estimate for $\nu$ and
hence it is plausible that
it satisfies the hyperscaling with $\nu=1/z=1/2$.

%%%%%%%%%%
\subsection{Effect of self doping}
\label{crossover}

Here, we discuss the effects of the level split
based on the numerical results obtained in Sec. \ref{levelcontrol}.
In this one-dimensional system,
our results indicate the crossover
between the Mott insulator with high spin state
and the band insulator with low spin state
with the self doping by the level split.

At $\varepsilon=0$, the system is in the Mott insulating state.
As discussed in Sec. \ref{NMIS},
this state has a finite spin gap and
a gapless excitation in the orbital channel.
In this state, two spins in degenerate orbitals couple with each other
ferromagnetically by the Hund coupling, that is,
they form the $S=1$ high spin state.
These $S=1$ spins show the short-range antiferromagnetic correlation
which leads to the peak structure of the spin correlation function
as shown in Fig. \ref{FIGSk}.

On the other hand, in the limit
$\varepsilon \gg U, J$ and $t$,
the system should behave as the band insulator.
Note that for $\varepsilon > 2t$,
the system becomes an insulator even in the non-interacting case.
In this limit, two orbitals are split by large $\varepsilon$
and particles predominantly occupy the one orbital $\nu=2$.
This behavior is clearly seen in Fig. \ref{FIGTzmom} (a).
For large values of $\varepsilon$,
a small density in the orbital $\nu=1$ may be induced by
the ${\cal H}_{J}$ term Eq. (\ref{dhmHJ}).
By the perturbation in $1/\varepsilon$,
the energy gain from this mixing is the order of
$-J^{2}/2\varepsilon$.
The fraction of the density in the orbital $\nu=1$ is estimated
by the $\varepsilon$ derivative of this gain,
therefore its leading term should be $O(\varepsilon^{-2})$.
As shown in Fig. \ref{FIGTzmom} (b),
the QMC data of $1-T_{\rm mom}^{z}$ for large $\varepsilon$
are well fitted by $a\varepsilon^{-2} + b\varepsilon^{-3}$.
If we assume the rigid-band picture,
the energy split between highest-occupied and lowest-unoccupied states
is given by $2\varepsilon-W^{*}$ with
the bandwidth $W^{*}$ renormalized by the interactions.
Then the $\varepsilon$ derivative of $-J^{2}/(2\varepsilon-W^{*})$
results in $1-T_{\rm mom}^{z} \simeq 
\frac{J^{2}}{2\varepsilon^{2}} \left(1+W^{*}/\varepsilon\right)$,
therefore, the ratio $b/a = W^{*}$.
From the fit in Fig. \ref{FIGTzmom} (b),
we obtain $W^{*}/t = 3.6 \pm 0.3$.
This indicates the bandwidth is narrowed by the interactions
compared with the non-interacting bandwidth $4t$
in this rigid-band scheme.
In the spin sector, in this band-insulator limit,
the spin moment is close to zero, that is,
the low spin state is realized.
This feature is clearly seen in the flat structure of
the spin correlation in Fig. \ref{FIGSk}.

Between these two limiting behaviors,
our QMC results obtained in Sec. \ref{levelcontrol}
suggest the crossover from one to the other.
Both the pseudo-spin moment and the spin correlation change continuously
by the level split, as shown in Figs. \ref{FIGTzmom} and \ref{FIGSk}.
In the region $\varepsilon \simle t$, as shown in Fig. \ref{FIGXk},
the anomalous correlation defined in Eq. (\ref{Xkdef})
has a peak structure at $k=\pi$.
This indicates that the Umklapp process
between different spins and orbitals is relevant to
the insulating nature in this region.
In the mean-field calculations,
the importance of this process is also pointed out
as discussed in Appendix C.
Of course, the mean-field description is known to break down
for one-dimensional systems:
It predicts the long-range order of this correlation and
the phase transition between the Mott insulator
and the band insulator as shown in Fig. \ref{eMFpd}.

When the degenerate orbitals split up
by increasing the value of $\varepsilon$,
our numerical results show that
the spin correlation is sensitive to a small self doping.
As shown in Fig. \ref{FIGSpivssd}, for $U=4J=3t$,
only $15\%$ self doping is enough to reduce
the value of $S(\pi)$ to the half of that at $\varepsilon=0$.
%For the $S=1$ Haldane system,
For spin systems with a finite spin gap,
the spin correlation function has
a Lorentzian-type peak at $k=\pi$ as
\begin{equation}
S(k) \propto \left[ \left(k-\pi\right)^{2}
                  + \xi_{\rm S}^{-2} \right]^{-1/2},
\end{equation}
where $\xi_{\rm S}$ is the spin correlation length.
This spin correlation length determines the spin gap;
$\Delta_{\rm S} \sim \xi_{\rm S}^{-1}$.
Although this form of $S(k)$ is not rigorously applicable to
the case for $\varepsilon \neq 0$,
it should be reasonable to consider that
the peak value of $S(k)$ at $k=\pi$
may be related to the spin correlation length as well as
the spin gap at least for $\varepsilon \ll U$.
The rapid decay of $S(\pi)$ against the self doping
in Fig. \ref{FIGSpivssd} suggests
the rapid growth of the spin gap from the value
in the Mott insulator at $\varepsilon=0$.
The inset of Fig. \ref{FIGSpivssd} indicates that
the spin gap increases linearly
with increasing the magnitude of self doping.

%%%%%%%%%%
\subsection{Comparison with other experimental and theoretical studies}
\label{remark}

In Sec. \ref{dhmdisc:scaling},
the universality class characterized by $\nu=1/z=1/2$
is suggested for the Mott transitions
controlled by the chemical potential
of the doubly degenerate Hubbard model in one dimension.
This result is considered to be relevant to
the metal-insulator transitions
when mobile holes are introduced into
the one-dimensional strongly correlated insulator
with a finite spin gap;
the $S=1$ Haldane system, the spin-Peierls system
or the spin ladder system with even-number legs.

Experimentally, the hole doping to the spin-gap material
has been realized in (Y,Ca)$_{2}$BaNiO$_{5}$
\cite{DiTusa1994,Kojima1995}.
The material Y$_{2}$BaNiO$_{5}$ is
a typical Haldane system with the spin gap $\sim 100K$
\cite{Cheong1992,Darriet1993},
where two spins in the $e_{g}$ orbital in the $d^{8}$ configuration
of Ni$^{2+}$ construct the one-dimensional network
as mentioned in Sec. \ref{ER}.
When we replace Y$^{3+}$ with Ca$^{2+}$ partially,
the charge doping into this chain is realized.
Unfortunately, in this system,
the doped holes did not show a clean metallic behavior,
although the density of states within the Haldane gap
is induced by them \cite{DiTusa1994}.
The temperature dependence of the magnetic susceptibility shows
a spin-glass like history at low temperature \cite{Kojima1995}.
These experimental results indicate that introduced holes
are localized by some disorder.
For the direct comparison with our result,
a material showing the metallic behavior
with hole doping is desired.
Effects of disorder on our theoretical result
also remains for further study for comparison
with the available experimental results.

Theoretically,
several one-dimensional models have been known to
have a finite spin excitation gap in the Mott insulating state;
for example,
the $t$-$J$-$J'$ model \cite{Ogata1991},
the dimerized $t$-$J$ model \cite{Imada1991},
the $t$-$J$ ladder model with even-number legs
\cite{Dagotto1996,Dagotto1992,Barnes1993,White1994}.
These models have been investigated with emphasis on
the possible persistence of the spin gap in the metallic phase
and on a realization of the superconductivity near the Mott insulator.
From this point of view,
the metallic region away from half filling
in our model provides another interesting example along this line.
However, in our model, another potential candidate of metals is
a ferromagnetic phase due to the double exchange mechanism.
In this work, we have not investigated this problem
mainly because of the negative sign difficulty
in numerical calculations as discussed in Appendix B.
This interesting problem remains for further study.

Next, we give remarks on the experimental relevance of the
crossover behavior indicated in Sec. \ref{crossover}.
As discussed in Sec. \ref{ER},
there are several one-dimensional systems
which are often called the Haldane materials.
In these systems,
the one-dimensional structure itself
may induce the deformation of the surrounding ligands,
which leads to
the level split between two orbitals in the $e_{g}$ state.
The magnitude of this level split is difficult
to estimate experimentally in general.
However,
it can be the same order as the bandwidth or the Hund coupling.
Our results suggest that
if this happens, the analysis based on
the theoretical results for the Heisenberg Hamiltonian
may not work straightforwardly
because of the gapless nature of orbital degrees of freedom
as is discussed in Sec. \ref{NMIS}.
%From the discussions in the previous subsection \ref{crossover},
%the value of the level split sensitively affects
%the spin gap and the spin state,
%and the nature of the band insulator is mixed continuously.
For the finite level split,
it is not clear whether the effective spin Hamiltonian
justified in the strong coupling limit is applicable.
Our result implies the necessity to start from
the tight-binding Hamiltonian as we discussed in this paper
at least for quantitative analysis.
%Therefore,
%we should be careful to discuss the effective spin Hamiltonian
%for these experimental situations.

In Sec. \ref{crossover}, the linear increase of the spin gap
with increase of the self doping is discussed.
As mentioned in Sec. \ref{ER},
the self doping may be induced by the change of
the ligand field, which is caused also by external pressure.
Recently, the Haldane material
Ni(C$_{2}$H$_{8}$N$_{8}$)$_{2}$NO$_{2}$(ClO$_{4}$) (NENP)
has been investigated under the external pressure
\cite{Ito1997}.
It shows that the spin gap grows
linearly with the pressure.
Since the pressure does not only simply increase
the level split or the self doping but also
the hopping matrix, that is, the bandwidth,
we need a care in analyzing it.
However, our result on the effect of self doping gives at least
one promising explanation for
the origin for the pressure dependence of the spin gap.

\section{Summary}
\label{Sec:Summary}

In this work, the multi-orbital Hubbard models are investigated
by the unbiased technique, the auxiliary field QMC method.
The general framework of the numerical method is introduced
for the multi-orbital models in detail.
This technique is applicable for arbitrary number of orbitals
in arbitrary dimension.
We present explicitly that there exist the non-trivial cases
where the negative sign problem is absent.
Using this new technique, the one-dimensional models
with doubly degenerate orbitals have been investigated in the ground state.
Some of basic properties of the Mott insulator at half filling are
quantitatively clarified.
The charge and spin gap amplitudes are calculated and
the characteristic behavior of them as the strongly correlated
insulator is clarified.
The charge gap can be one order of magnitude larger
than the spin gap in our result of the Mott insulating phase.
The orbital degrees of freedom are shown to have a gapless excitation.
By controlling the chemical potential,
the critical properties of the insulator-metal transition
have been investigated.
We obtain the correlation length exponent $\nu=1/2$ which is consistent
with the general prediction by the scaling arguments.
The universality class for this transition has been discussed
by analyzing our numerical results based on the hyperscaling hypothesis.
For the level split between two orbitals,
the effect of self doping is examined in detail.
Our numerical results of correlation functions indicate
the crossover between the Mott insulator with high spin state
and the band insulator with low spin state.
The significance of the Umklapp scattering
between different spins and orbitals in this crossover is suggested.
With increase of the level split,
remarkably different spin and charge gap amplitudes
in the Mott insulating phase continuously crossover
to a similar value in the band insulating phase.
The antiferromagnetic spin correlation as well as
the inverse of the spin gap
sensitively decreases with the self doping.
Our result of the chemical potential control may have relevance to
the hole doping of the Haldane materials.
Our result of the rapid increase of the spin gap
by the self doping is consistent with
the experimentally observed increase of
the spin gap in the Haldane materials under pressure.

%%%%%%%%%%
\section*{Acknowledgement}

The authors appreciate Nobuo Furukawa 
for stimulating discussions and important suggestions.
They also thank Nobuyuki Katoh and Takeo Kato
for fruitful discussions.
This work is supported
by `Reseach for the Future Program from Japan Society
for Promotion of Science (JSPS-RFTF 97P01103) as well as
by a Grant-in-Aid for Scientific Research
on the Priority Area `Anomalous Metallic State near the Mott Transition'
from the Ministry of Education, Science, Sports and Culture, Japan.
A part of the computations in this work was performed
using the facilities of the Supercomputer Center,
Institute for Solid State Physics, University of Tokyo.

%%%%%%%%%%
\section*{Appendix A}

In this part, numerical convergence in the actual QMC calculations
is examined in detail.
As described in Sec. \ref{Sec:Method},
in the auxiliary field algorithm for the ground state,
we should take care of the convergence about two quantities;
one is the width of the Suzuki-Trotter slice $\Delta\tau$ and
the other is the magnitude of the projection $\beta$.
We discuss the behavior of these systematic errors
for the doubly degenerate model defined in Sec. \ref{DDM}.

\begin{figure}
\hfil
\epsfxsize=8cm
\hfil
\epsfbox{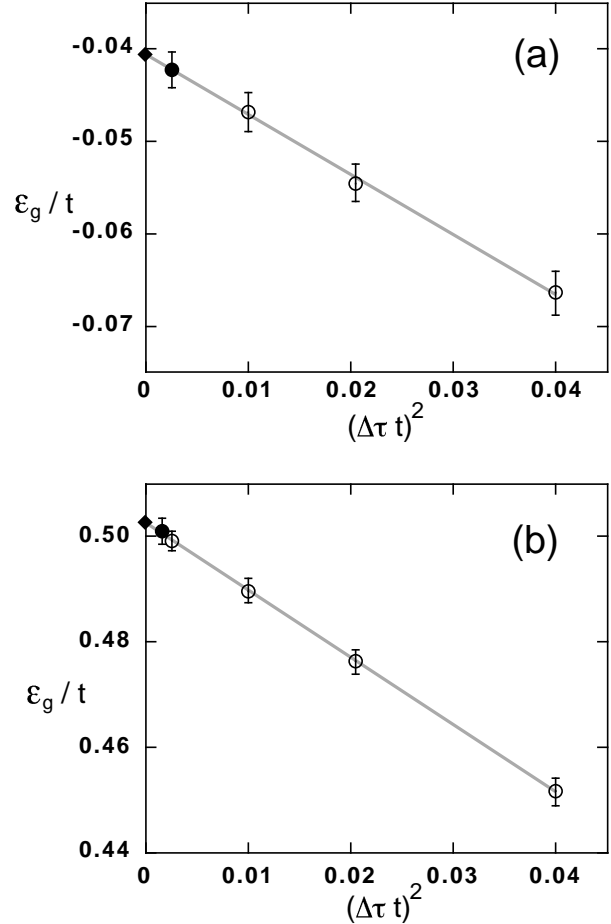}
\caption{
$\Delta\tau$ dependence of the QMC data for the ground state energy
of the doubly degenerate Hubbard model.
The data are for $14$-site systems with $\varepsilon=0$;
(a) $U=4J=2t$ and (b) $U=4J=3t$.
Extrapolated values 
to $\Delta\tau \rightarrow 0$ by the fit shown as the gray lines
are plotted by the diamonds.
Filled circles are the data for the values of $\Delta\tau$
which we employ in the actual calculations.
See text for details.
}
\label{FIG:appEgExt}
\end{figure}

First, we examine the convergence about $\Delta\tau$.
As discussed in Sec. \ref{Sec:Method},
the systematic errors dominantly come from
the Suzuki-Trotter decomposition Eq. (\ref{STdecomp}).
The magnitude of the error depends on the parameters in the Hamiltonian
as well as on the physical quantities which we measure.
Fig. \ref{FIG:appEgExt} shows the $\Delta\tau$ dependence
of the ground state energy per site and orbital,
$\varepsilon_{\rm g} \equiv E_{\rm g} / 2 N_{\rm S}$,
for two sets of parameters,
(a) $U=4J=2t$ and (b) $U=4J=3t$.
The data are for $14$-site systems with $\varepsilon=0$.
Among the quantities which we have calculated,
the ground state energy has the strongest dependence on
$\Delta\tau$.
As shown in Fig. \ref{FIG:appEgExt},
the QMC data are well fitted by 
\begin{equation}
\varepsilon_{\rm g} = \varepsilon_{\rm g}^{\rm ext}
- a \left(\Delta\tau\right)^{2},
\end{equation}
as expected from the decomposition Eq. (\ref{STdecomp}).
Here $\varepsilon_{\rm g}^{\rm ext}$ denotes the extrapolated value
to the limit $\Delta\tau\rightarrow 0$,
which is plotted as the diamonds in Fig. \ref{FIG:appEgExt}.
In actual QMC calculations,
instead of obtaining $\varepsilon_{\rm g}^{\rm ext}$ by this extrapolation,
we take small $\Delta\tau$ enough to guarantee the convergence
within numerical errorbars for each interaction parameters.
For example, we employ
$\Delta\tau t= 0.05$ for $U=4J=2t$ and
$\Delta\tau t= 0.04$ for $U=4J=3t$.
From Fig. \ref{FIG:appEgExt},
these values of $\Delta\tau$ are found to ensure
the numerical convergence to the limit $\Delta\tau\rightarrow 0$.

\begin{figure}
\hfil
\epsfxsize=8cm
\hfil
\epsfbox{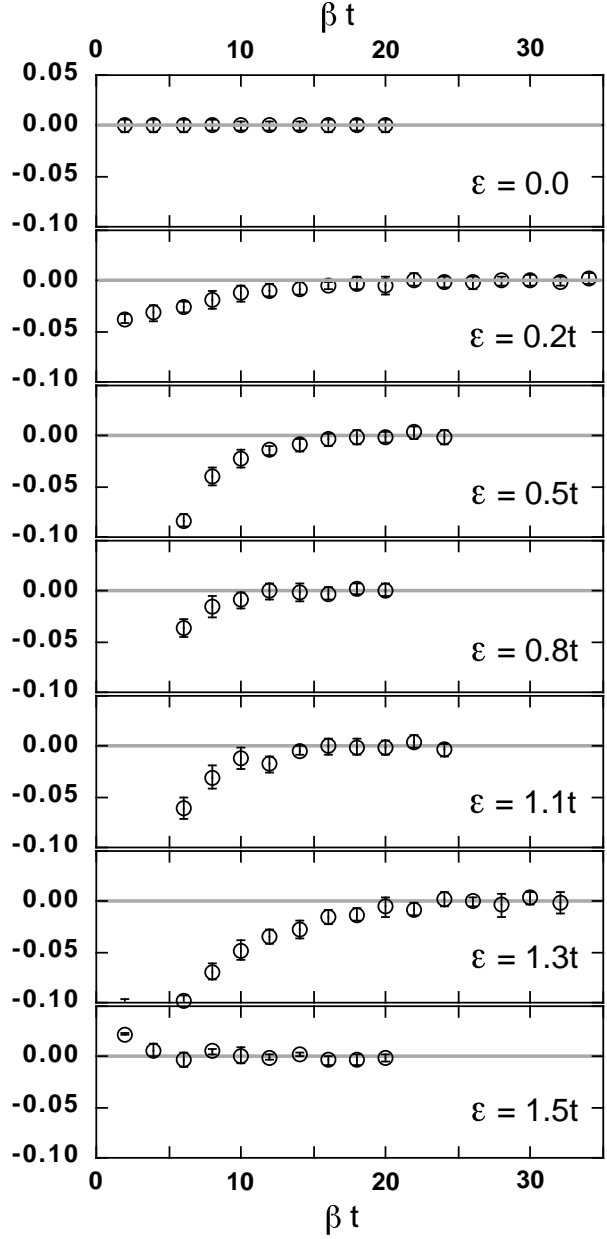}
\caption{
$\beta$ dependence of the QMC data for the pseudo-spin moment
of doubly degenerate Hubbard model
for various values of the level split.
The data are for $14$ site systems with $U=4J=3t$.
Differences from the converged values which we determine are plotted.
}
\label{FIG:tauExt}
\end{figure}

Next, the convergence about the projection $\beta$ is discussed.
In general, the ground state should be projected out
from a trial state for large $\beta$
compared with the inverse of the smallest energy scale in the system.
Therefore, the values of $\beta$ sufficient to obtain converged data
strongly depend on the model parameters
as well as the trial state.
We illustrate this behavior for
the change of the level split $\varepsilon$.
Fig. \ref{FIG:tauExt} shows the $\beta$ dependence
of the pseudo-spin moment defined in Eq. (\ref{Tzmom})
for various values of $\varepsilon$.
Deviations from the converged values
are plotted in the figures.
%We find relatively slow convergence
%where the self doping occurs conspicuously.
%See also Appendix 5.B.
In actual calculations, we check
the convergence for each parameter and physical quantity carefully,
and obtain the results by averaging
several data for different values of $\beta$
in the converged region.

%%%%%%%%%%
\section*{Appendix B}

Here we discuss the behavior of the negative sign
in our framework introduced in Sec. \ref{Sec:Method}.
Especially, we focus two different elements which break
the particle-hole symmetry and lead to the negative sign problem;
the long-range hopping and the hole or electron doping.

\begin{figure}
\hfil
\epsfxsize=8cm
\hfil
\epsfbox{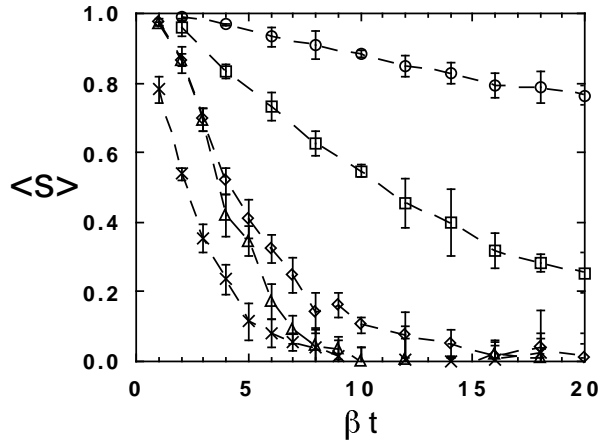}
\caption{
$\beta$ dependence of the negative sign for
the change of the next-nearest neighbor hopping
in $10$ site systems with $U=4J=2t$.
The dashed lines are guides to eyes.
The data are for $t'/t = 0.2, 0.4, 0.6, 0.8$ and $1.0$
from the top to the bottom.
}
\label{appBsign1}
\end{figure}

First, we examine the effect of the long-range hopping.
We add the next-nearest neighbor hopping proportional to $t'$
to the doubly degenerate model Eqs. (\ref{dhmHt}) $\sim$ (\ref{dhmHJ}).
Note that this term is different from
the next-nearest neighbor hopping term 
discussed in Sec. \ref{Sec:Conditions};
here we consider
$t_{\langle\langle ij \rangle\rangle}^{11} =
 t_{\langle\langle ij \rangle\rangle}^{22} = t' \ge 0$.
Fig. \ref{appBsign1} shows the behavior of the negative sign
when we change the value of $t'$ at half filling.
The negative sign $\langle S \rangle$ is defined as
the real part of the phase of MC weights,
\begin{equation}
\langle S \rangle = \Re 
\frac{\langle W_{\uparrow} W_{\downarrow} \rangle}
     {\langle | W_{\uparrow} W_{\downarrow} | \rangle}.
\end{equation}

\begin{figure}
\hfil
\epsfxsize=8cm
\hfil
\epsfbox{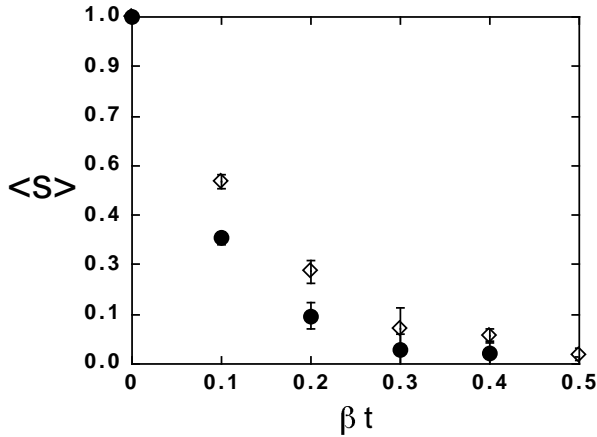}
\caption{
$\beta$ dependence of the negative sign for the hole doping.
The filled circles are for
$N_{\rm S} = 6$ and $N_{\rm e} = 20$.
The open diamonds are for
$N_{\rm S} = 10$ and $N_{\rm e} = 12$.
}
\label{appBsign2}
\end{figure}

Next, the effect of the hole doping is discussed.
Fig. \ref{appBsign2} plots the values of $\langle S \rangle$
when we introduce the extra holes or electrons to the model
Eqs. (\ref{dhmHt}) $\sim$ (\ref{dhmHJ}) at half filling.
We choose the electron filling which is closed-shell configuration.
As shown in Fig. \ref{appBsign2},
the negative sign difficulty is far severe in this case as
compared with the results in Fig. \ref{appBsign1}.
This may be partly because our HS transformation Eq. (\ref{expU})
contains the total density as the phase factor;
The change of the electron filling seems to be crucial
to break the phase coherence of each MC samples
which is maintained at half filling.

%%%%%%%%%%
\section*{Appendix C}

In this Appendix,
by using the mean-field approximation,
we discuss overall features of phase diagrams
for the doubly degenerate Hubbard model
defined in Eqs. (\ref{dhmHt}) $\sim$ (\ref{dhmHJ}).
In Sec. \ref{Sec:Results},
the QMC calculations are done for one-dimensional systems.
It is difficult to compare the mean-field results
with the QMC results directly,
because strong quantum fluctuations in one dimension
lead to the breakdown of the mean-field approach.
Nevertheless, the mean-field picture is useful
to understand the basic properties of the model.

For the chemical potential control,
the mean-field phase diagram
for the doubly degenerate Hubbard model has been investigated
by Inagaki and Kubo in three dimensions \cite{Inagaki1973}.
They have especially concentrated on various spin and orbital states.
We discuss here their results near half filling
which are the issue of our investigation
in Sec. \ref{Sec:Results} and \ref{Sec:Discussion}.
For $\varepsilon=0$, at half filling,
the system becomes an insulator with spin antiferromagnetic
long-range order and no orbital ordering
for infinitesimal Coulomb interaction.
When we change the chemical potential $\mu$,
the insulator-metal transition occurs
and
the metallic state with the antiferromagnetic ordering appears.
For further change of $\mu$,
this spin ordering disappears and the metal
without any ordering of spin and orbital components is realized.
All these features are common to
the mean-field phase diagram for the single-orbital Hubbard model
\cite{Penn1966}.

For the change of
%the next-nearest neighbor hopping $t'$ or
the level split $\varepsilon$,
we obtain the Hartree-Fock phase diagrams
following the method by Inagaki and Kubo \cite{Inagaki1973}.
Here we fix the interaction ratio $U/J=4$.
Our Hartree-Fock calculations are done in one dimension.

\begin{figure}
\hfil
\epsfxsize=8cm
\hfil
\epsfbox{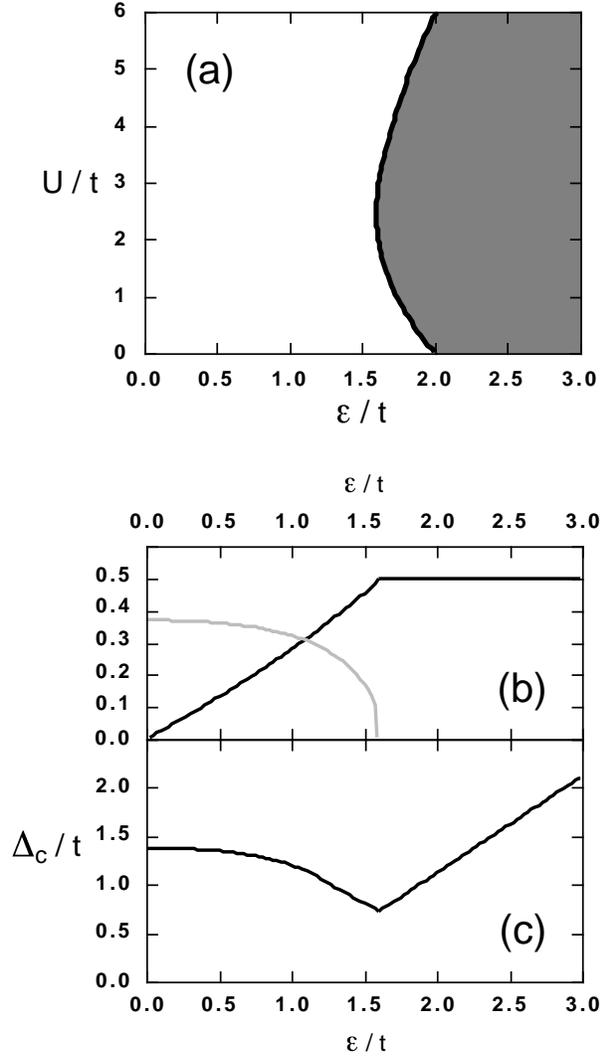}
\caption{
Mean-field results for the level split control;
(a) Phase diagram. We fix $U=4J$.
The gray area represents a band insulator.
The white area shows an insulating state
due to the nesting properties between different spins and orbitals.
(b) The mean fields as function of $\varepsilon$ at $U=4J=2t$.
The black curve denotes 
$(-1)^{\nu} \left( \langle n_{i\nu\sigma} \rangle -1/2 \right)
= \langle T_{\rm mom}^{z} \rangle /2$,
which corresponds to the magnitude of the self doping.
The gray one is for 
$\langle c_{k\nu\sigma}^{\dagger} c_{k+\pi,\nu'\sigma'} \rangle$.
(c) $\varepsilon$ dependence of the charge gap at $U=4J=2t$.
}
\label{eMFpd}
\end{figure}

Fig. \ref{eMFpd} (a) shows the mean-field phase diagram
in the plane of the level split $\varepsilon$ and
the interaction strength $U$.
At $\varepsilon=0$, the antiferromagnetic insulator appears
for $U > 0$ as mentioned above.
In this state, the antiferromagnetic order parameter
$\langle c_{k\nu\uparrow}^{\dagger} c_{k+\pi\nu\downarrow} \rangle$
has the same value as
$\langle c_{k\nu\uparrow}^{\dagger} c_{k+\pi\nu'\downarrow} \rangle$
because of the symmetry about the orbital index $\nu$
at $\varepsilon=0$.
When we switch on $\varepsilon$, the latter
$\langle c_{k\nu\uparrow}^{\dagger} c_{k+\pi\nu'\downarrow} \rangle$
survives and the self doping begins
as shown in Fig. \ref{eMFpd} (b).
There, the insulating state persists for $\varepsilon \neq 0$
presumably because of the nesting properties
between different spins and orbitals
as discussed in the last part of Sec. \ref{Sec:Method}.
The self doping occurs continuously
until all the electrons occupy the orbital $\nu=2$, that is,
the phase transition to the band insulating state.
The mean-field results for the charge gap $\Delta_{\rm C}$
which is defined in Eq. (\ref{chgapdef0})
are plotted in Fig. \ref{eMFpd} (c).
The charge gap has a finite value for all the values of $\varepsilon$.
However, the origin of the gap is clearly different
between two phases in Fig. \ref{eMFpd} (a).
For small values of $\varepsilon$,
the gap originates from the nesting properties and the ordering of
$\langle c_{k\nu\uparrow}^{\dagger} c_{k+\pi\nu'\downarrow} \rangle$,
and the value of the gap decreases gradually.
Compared to this, for large values of $\varepsilon$,
the gap increases linearly as the function of $\varepsilon$,
which is expected for the band insulating state.

\end{document}